\documentclass[11pt]{article}

\usepackage[title]{appendix}

\usepackage{fancyvrb}

\usepackage{caption}

\usepackage{listings}

\usepackage{xfrac}

\usepackage{pgfplots}
\pgfplotsset{width=1.0\textwidth,compat=1.9}

\usepackage{empheq}
\usepackage[shortlabels]{enumitem}
\setlist[enumerate]{nosep}
\usepackage[doc]{optional}
\usepackage{xcolor}
\usepackage[colorlinks=true,
            linkcolor=refkey,
            urlcolor=lblue,
            citecolor=red]{hyperref}
\usepackage{float}
\usepackage{soul}
\usepackage{graphicx}
\definecolor{labelkey}{rgb}{0,0.08,0.45}
\definecolor{refkey}{rgb}{0,0.6,0.0}
\definecolor{Brown}{rgb}{0.45,0.0,0.05}
\definecolor{lime}{rgb}{0.00,0.8,0.0}
\definecolor{lblue}{rgb}{0.5,0.5,0.99}

 \usepackage{mathpazo}

\usepackage{stmaryrd}
\usepackage{amssymb}

\usepackage{booktabs}

\newcommand{\seppfive}{\setlength{\itemsep}{-5pt}}

\newcommand{\fenv}[1]%
{\ensuremath{\,\overrightarrow{\operatorname{env}}_{#1}}}
\newcommand{\benv}[1]%
{\ensuremath{\,\overleftarrow{\operatorname{env}}_{#1}}}

\newcommand{\sthree}{\ensuremath{\operatorname{s3}}}

{\begin{list}{}{%
\settowidth{\labelwidth}{\textrm{#1~}}%
\setlength{\leftmargin}{\labelwidth+\labelsep}}}
{\end{list}}
\usepackage{amsthm}
\usepackage[capitalize,nameinlink]{cleveref}
\crefname{equation}{}{equations}
\crefname{chapter}{Appendix}{chapters}
\crefname{item}{}{items}
\crefname{enumi}{}{}
\newtheorem{theorem}{Theorem}[section]

\newtheorem{ex}[theorem]{Example}

\definecolor{myblue}{rgb}{.8, .8, 1}

\allowdisplaybreaks 

\begin{document}

\author{
Alan W.\ Paeth\thanks{
\emph{(formerly)} Computer Science, University
of British Columbia,
Kelowna, B.C.\ V1V~1V7, Canada.
}
}

\title{ \textsc
Non-Interfering Concurrent Exchange (NICE) Networks
}

\date{August 9, 2018}

\maketitle

\begin{abstract}

\noindent
In studying the statistical frequency of exchange in comparison-exchange (CE)
networks we discover a new elementary form of comparison-exchange which we name
the "2-op". The operation supports concurrent and non-interfering operations of
two traditional CEs upon one shared element. More than merely improving overall
statistical performance, the introduction of NICE (non-interfering CE) networks
lowers long-held bounds in the number of stages required for sorting tasks.
Code-based CEs also benefit from improved average/worst case run time costs.
\end{abstract}
{
\noindent

}

\section*{Preamble}
This manuscript is a refined version of one of the last projects
Dr.~Alan W.~Paeth, a Professor of Computer Science at the University of
British Columbia Okanagan worked on. Sadly, Dr.~Paeth was not able to
complete his work due to his courageous battle with cancer \cite{Obituary}.

We (Heinz Bauschke, Scott Fazackerley, Wade Klaver, Mason Macklem) have taken up
Dr.~Paeth's last draft (see Appendix~\ref{app:AlanEmail}) and attempted
to polish it, to
connect it to existing literature (see \cite[Section~5.3.4]{Knuth}),
and to disseminate its contents.
Please contact us at \url{heinz.bauschke@ubc.ca} for questions and comments
concerning this manuscript, or Dr.~Paeth's son Doug at 
\url{dpaeth@gmail.com}.

\section{Motivation}

Comparison-based sorts dominate much of modern sorting; in-place
methods such as quicksort are widely employed and well-studied. All
seek to minimize the number of comparisons required. For small
numbers of input ($N$),
a sequence of predetermined comparisons form
a decision tree of $N!$ leaves;
if minimal, its height is then
$\lceil N\log_{2} N\rceil$
e.g. for $N=5$, $7$
comparisons suffice to fully determine the
input permutation.
To complete the sort a sequence of cyclic exchanges
then reorder the data.
These are often simplified into a sequence
of two-element swaps.
At $N>5$, the decision tree's size makes it
impractical in production settings.
At $N=3$ all solutions require a
tree having $\texttt{height}=3$, $\texttt{leaves}=6$.
The tree show below is optimal in
that the two comparison descents occur with
$\texttt{mem}[1]\leq \texttt{mem}[2]\leq \texttt{mem}[3]$
or
$\texttt{mem}[1]\geq \texttt{mem}[2]\geq \texttt{mem}[3]$ i.e.,
termination occurs early with ascendingly presorted
elements:

\begin{verbatim}
        if mem[1] <= mem[2] then
             if mem[2] <= mem[3]      then      return // 1 2 3
             else if mem[1] <= mem[3] then      return // 1 3 2
                  else                          return // 3 1 2
        else if mem[2] >= mem[3]      then      return // 3 2 1
             else if mem[1] >= mem[3] then      return // 2 3 1
                  else                          return // 2 1 3
\end{verbatim}

This network is well known.
In the first two steps we either establish
\verb+mem[1]<=mem[2]+ and
\verb+mem[2]<=mem[3]+ and gain
\verb+mem[1]<=mem[3]+ by transitivity
(we are sorted).
Otherwise, at the first
\verb+else+,
\verb+mem[2]+
has largest rank; we need merely
disambiguate the ranks of nonadjacent
\verb+mem[1]+ and
\verb+mem[3]+.
The second half follows by symmetry.

(For larger $N$,
we can establish that
\verb-mem[i]<mem[i+1]- in $N-1$ steps
but this set of
comparisons does not lead to an optimal
(balanced) decision tree.)
In the code seen above,
exchanges complete the sort.
For the above six leaves, these
are the permutations (written in cycle notation)
\texttt{nil},
(23),
(123),
(13),
(132),
(12),
respectively.

By contrast,
compare-exchange sorting networks order an array
by performing a fused compare and
conditional-exchange operation.
(See, e.g., \cite[Section~8.6]{AHU} and
\cite[Section~5.3.4]{Knuth} for further information.)
They are ideally suited to sorting arrays of integers of
fixed small size ($N = 1\ldots 25$ typical)
where the decision making overhead of more
general methods (tree traversal using comparison and bifurcation)
will diminish or even negate any reduction in total machine
comparisons. An integer compare is typically a single machine
instruction. While compact and efficient,
a general methodology for
the creation of optimal fixed compare-exchange networks
remains elusive.

\begin{figure}[H]
\begin{center}
\begin{tabular}{c c c}
\includegraphics[scale=0.15]{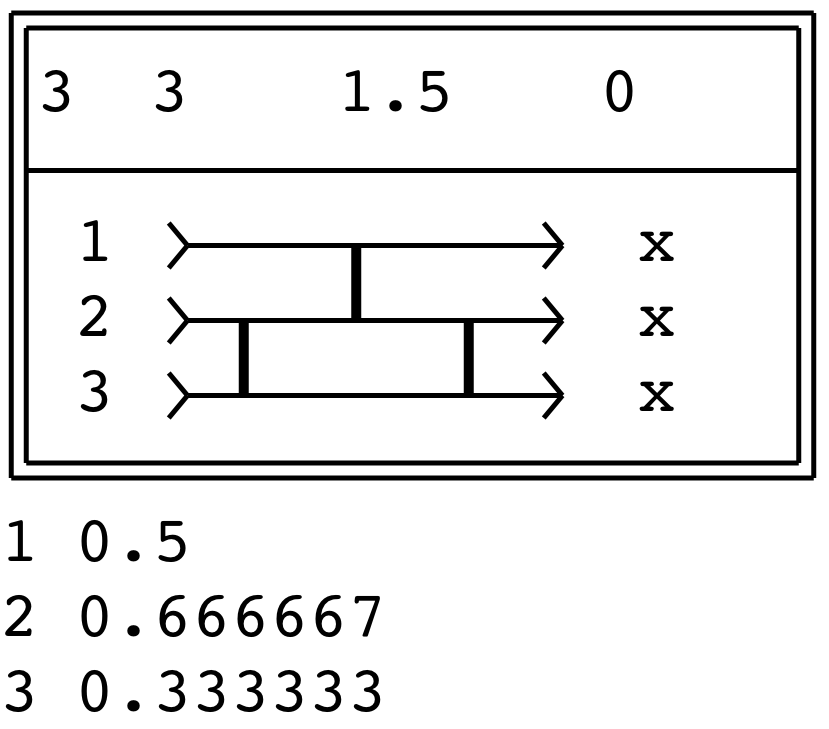}
& &
\includegraphics[scale=0.15]{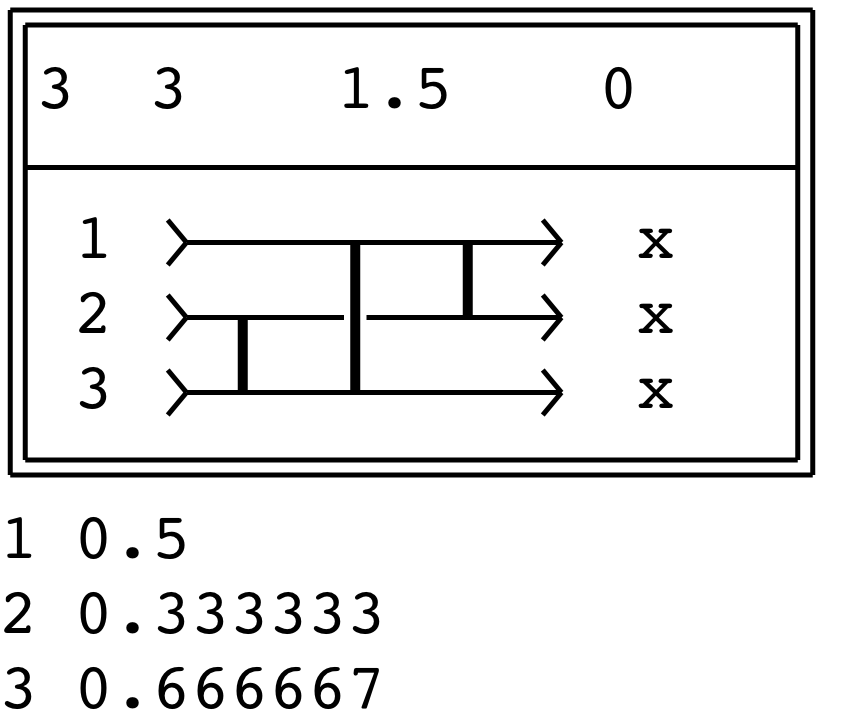}    \\
(a) & & (b)
\end{tabular}
\end{center}
 \caption{\small
 Two sorting networks:
 (a) corresponds to the comparisons
 (23), (12), (23), at stages 1, 2, 3, respectively;
 (b) corresponds to the comparisons
 (23), (13), (12), at stages 1, 2, 3, respectively.
  }
\label{fig:1}
\end{figure}

Worst,
lack of suitable metrics may lead to sub-optimal networks.
Below are two simple networks for $N=3$.
In Figure~\ref{fig:1}(a), we recode based on the above algorithm.
In Figure~\ref{fig:1}(b), we apply
Batcher's even-odd construction
for an odd number of elements,
applying central symmetry (of inversion)
to the right and left halves.

Exchanges are costly, often at a ratio of
$2:1$ or $3:1$ to a simple comparison.
We have affixed the likelihood of an exchange to each link.
Summary statistics
give the maximum and average number of exchanges for the entire network.
Consider Figure~\ref{fig:1}(a) which features
3 stages, namely ---in that order--- (23), (12), and again (23).
The number \texttt{3} in the upper left corner gives the \emph{number}
of links for this particular network.
The second number, again \texttt{3} here, means that there is
one input to the network where $3$ links are active and swap,
the \emph{worst-case scenario}.
The third number, \texttt{1.5}, means that the average number of swaps,
taken over all possible inputs, is $1.5$.
The number \texttt{0} in the upper right corner is a
degree of \emph{disorder} meaning that this design results in a
fully sorted array which can also be seen by the three \texttt{x}
characters in the output.
Below the left rectangle,
the first row ``\verb+1 0.5+'' signifies that the probability of an exchange at
stage 1 is $0.5=\tfrac{1}{2}$;
the second row ``\verb+2 0.666667+'' signifies that the probability of an exchange at
stage 2 is $0.666667 \approx 2/3$;
finally, the third row
``\verb+3 0.333333+'' signifies that the probability of an exchange at
stage 3 is $0.333333 \approx 1/3$.
In Appendix~\ref{app:details},
we provide more details on how the statistics in
Figure~\ref{fig:1}(a) was obtained.

The networks presented so far for $N=3$ are
distinct and require in both cases
three stages, comparisons and exchanges.
Both also demonstrate that a full
sort network will include all possible $\texttt{length}=1$
links as these alone can serve
to reorder permutations of sorted data when
merely one adjacent transposition exists.
The symmetry in Figure~\ref{fig:1}(b) is compelling:
it allows for fully bidirectional
sorting when then input and output sides reverse.
Unfortunately, it (as in Figure~\ref{fig:1}(a)) also
exhibits a link (a.k.a.\ comparator in \cite{Knuth})
in which swapping occurs more often than not.

When a link exchange occurs more than half the time,
we may reverse the swap vs
non-swap $.67/.33$ to $.33/.67$
by preexchanging its elements.
We remove the direct cost of the exchange by reversing
the order of input lines to the left of the
offending link. But all inputs are unsorted so
any non-conditional exchange has
no sorting efficacy and may be removed.
In Figure~\ref{fig:1}(a),
the offending link is at stage 2, namely (12).
Thus, we replace (23) by (13) at stage~1
while keeping (23) at stage~3 unchanged.
For Figure~\ref{fig:1}(b),
the offending link is at stage~3, namely (12).
Thus, we replace (23) by (13) at stage~1
and (13) by (23) at stage~2.

\begin{figure}[H]
\begin{center}
\begin{tabular}{c c c}
\includegraphics[scale=0.15]{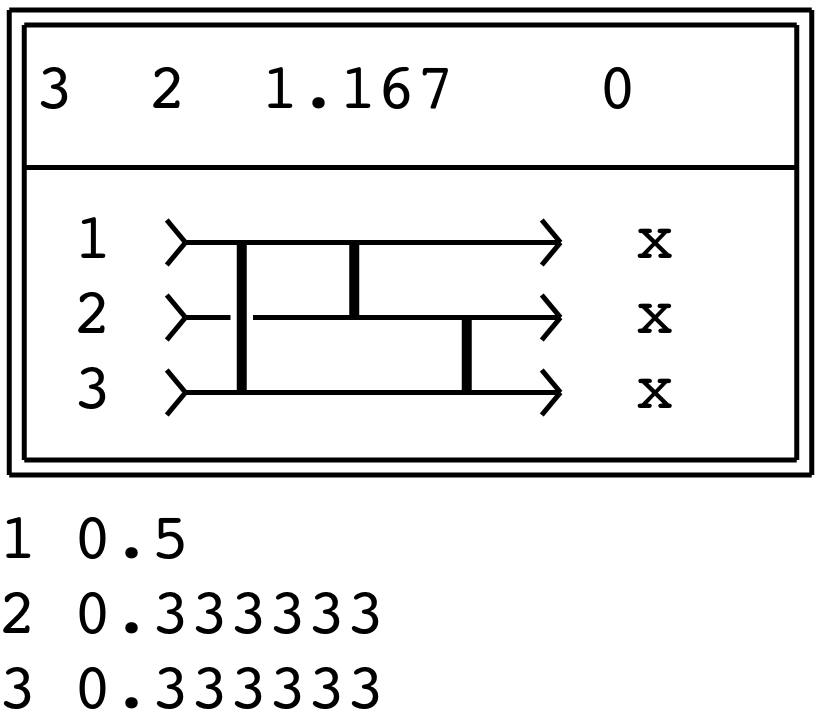}
& &
\includegraphics[scale=0.15]{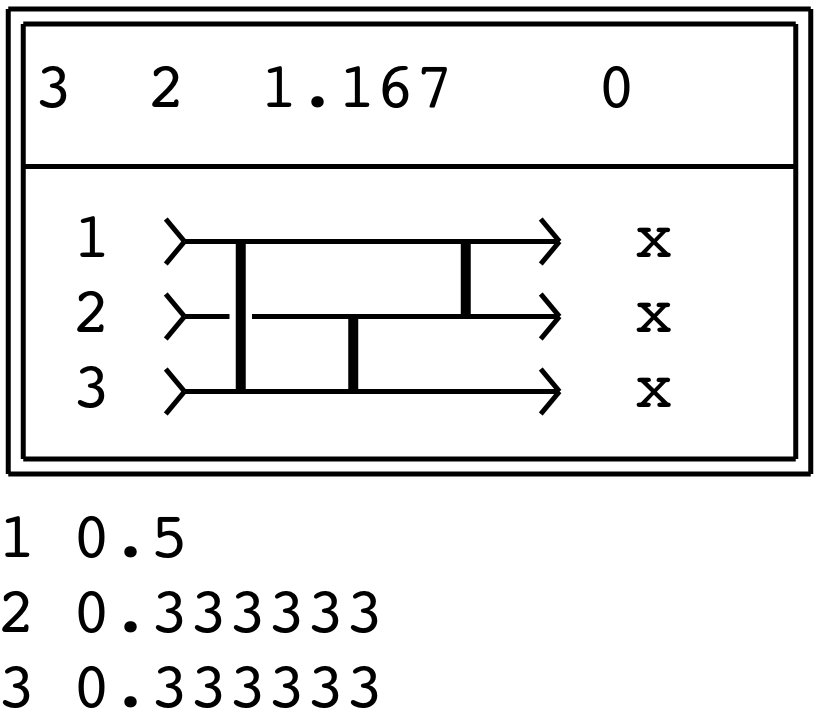}    \\
(a) & & (b)
\end{tabular}
\end{center}

 \caption{\small
 Two sorting networks:
 (a) corresponds to the comparisons
 (13), (12), (23), at stages 1, 2, 3, respectively;
 (b) corresponds to the comparisons
 (13), (23), (12), at stages 1, 2, 3, respectively.
}
\label{fig:2}
\end{figure}
The networks, depicted in Figure~\ref{fig:2} and
identical under mirror symmetry,
substantially improve the
average cost of exchange.
More striking is a reduction in maximum exchange,
which was lowered from 3 to 2!
Moreover, the average number of swaps was lowered from
$3/2=1.5$ to $7/6\approx 1.167$.
This is unexpected and serves as the basis
of non-interfering concurrent exchange.
Clearly, the first link may exchange
without restriction.
The remaining cost of one exchange must be shared between
the two remaining links: \emph{at most one} may occur.
This in turn implies that the
central element cannot be rewritten by both links,
allowing both the upper and
lower link concurrent execution.

We now create a new fused dual-link element,
which we name a ``\emph{2-op}'' (borrowing
from mathematical nomenclature used for such occurrences), depicted below:

\begin{figure}[H]
\begin{center}
\includegraphics[scale=0.15]{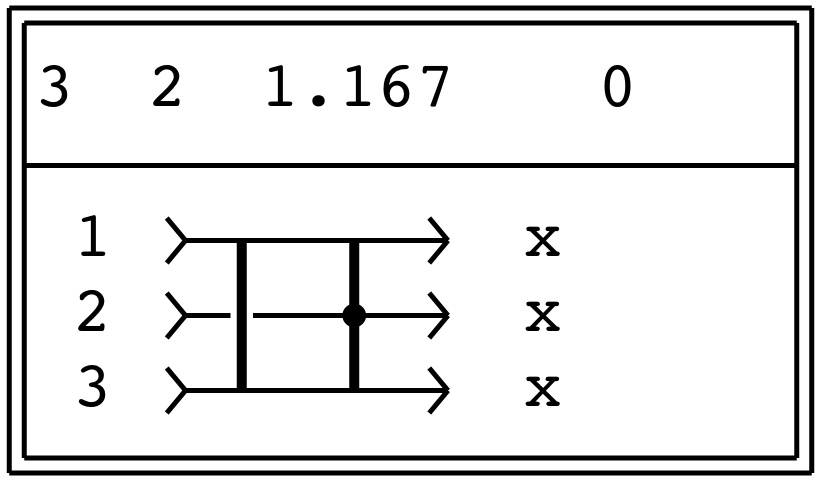}
\end{center}
 \caption{\small
At stage~1, we have the comparison $(13)$,
while at stage~2, we have the new 2-op $(123)$
which will execute at most one of the exchanges
$\{(12),(23)\}$.
}
\label{fig:3}
\end{figure}
The circle at the common link joint $2$ indicates non-interference.

\section{9-element networks and medians}

The 2-op exchange on three elements yields immediate gains in
lowering costs of traditional networks.
For example, Paeth \cite{Paeth90} described a median on a
$3\times 3$
box (see Figure~\ref{fig:4}),
\begin{figure}[H]
\centering
\begin{tabular}[t]{|c|c|c|}
\hline
1 & 2 & 3 \\
\hline
4 & 5 & 6 \\
\hline
7 & 8 & 9 \\
\hline
\end{tabular}
\caption{Labeling of cells in the $3\times 3$ box.}
\label{fig:4}
\end{figure}%
\noindent
formed by column, row and (single) diagonal sorting, first conceived as
a means to reuse column sorts when filtering a large raster image.

Defining $\sthree(a,b,c)$ to be either the three swaps
$(a,b)(b,c)(a,b)$ (as in the original \cite{Paeth90})
or $(a,c)(a,b)(b,c)$ (reworked to be a 2-op!),
we can find the median by
$\sthree(1,4,7)$,
$\sthree(2,5,8)$,
$\sthree(3,6,9)$,
$\sthree(1,2,3)$,
$\sthree(4,5,6)$,
$\sthree(7,8,9)$,
$\sthree(3,5,7)$,
where this sequence of operators  is executed from left to right.

\subsection{Original design}

The first version gives the 21-link network in Figure~\ref{fig:5}
below.

\begin{figure}[H]
\begin{center}
\includegraphics[width=0.8\textwidth]{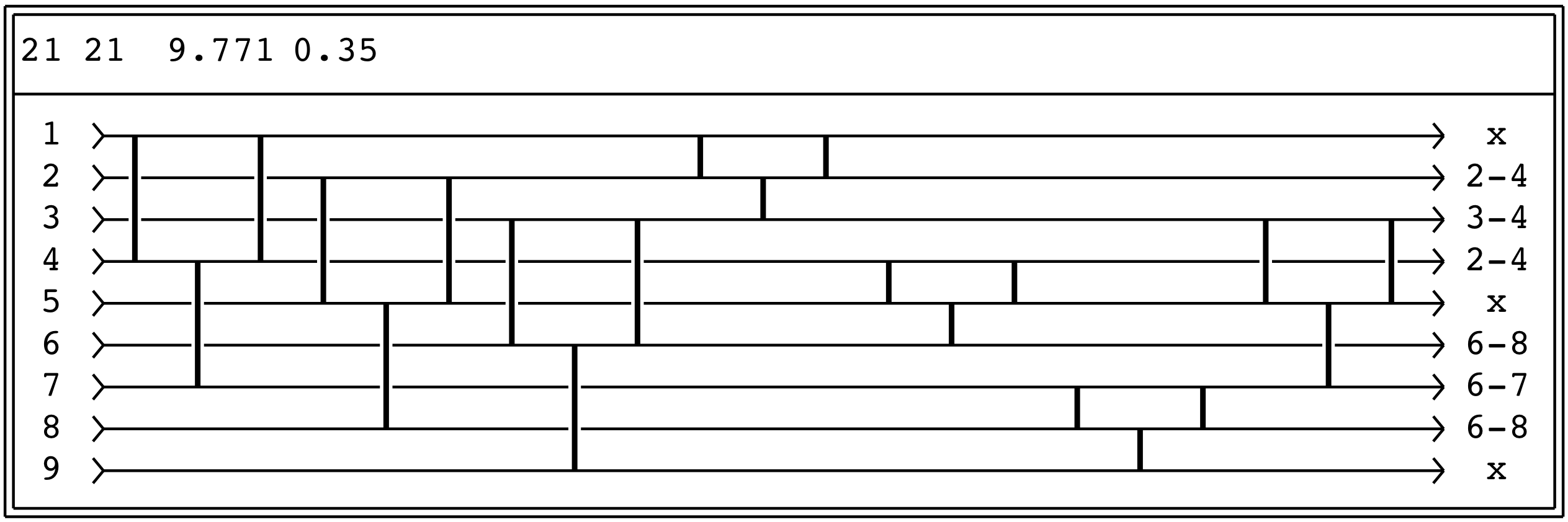}
\end{center}
 \caption{\small
 A 21-link network to determine the median.
The probabilities of swapping are:
$\sfrac{1}{2}$ at link~1,
$\sfrac{2}{3}$ at link~2,
$\sfrac{1}{3}$ at link~3,
$\sfrac{1}{2}$ at link~4,
$\sfrac{2}{3}$ at link~5,
$\sfrac{1}{3}$ at link~6,
$\sfrac{1}{2}$ at link~7,
$\sfrac{2}{3}$ at link~8,
$\sfrac{1}{3}$ at link~9,
$\sfrac{1}{2}$ at link~10,
$\sfrac{2}{3}$ at link~11,
$\sfrac{1}{3}$ at link~12,
$\sfrac{1}{2}$ at link~13,
$\sfrac{2}{3}$ at link~14,
$\sfrac{1}{3}$ at link~15,
$\sfrac{1}{2}$ at link~16,
$\sfrac{2}{3}$ at link~17,
$\sfrac{1}{3}$ at link~18,
$\sfrac{19}{70}$ at link~19,
$\sfrac{5}{14}$ at link~20,
$\sfrac{1}{7}$ at link~21.
This results in 
$9\sfrac{27}{35}\approx 9.771$
swaps on average.
}
\label{fig:5}
\end{figure}

In Figure~\ref{fig:180619}, we re-arrange the links in
Figure~\ref{fig:5}
to highlight \emph{stages}, i.e., links that can be executed in
parallel.
(Stages correspond to delay time in \cite{Knuth}.)
This does not change the overall swap statistics:

\begin{figure}[H]
\begin{center}
\includegraphics[width=0.8\textwidth]{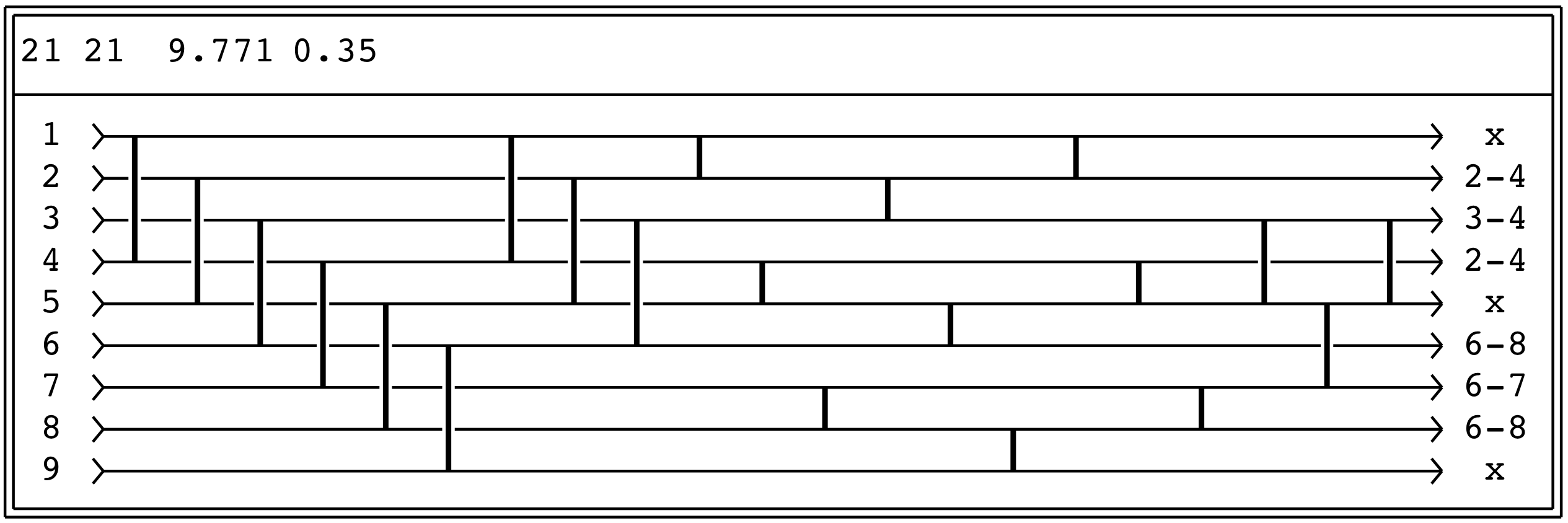}
\end{center}
 \caption{\small
The 21-link network to determine the median from
Figure~\ref{fig:5} re-arranged.
Note that there are 9 stages
(stage~1: links $(1,4)(2,5)(3,6)$;
stage~2: links $(4,7)(5,8)(6,9)$;
stage~3: links $(1,4)(2,5)(3,6)$;
stage~4: links $(1,2)(4,5)(7,9)$;
stage~5: links $(2,3)(5,6)(8,9)$;
stage~6: links $(1,2)(4,5)(7,9)$;
stage~7: link $(3,5)$;
stage~8: link $(5,7)$;
stage~9: link $(3,5)$)
that can be executed in parallel.
}
\label{fig:180619}
\end{figure}

Using 21 links, we find minimum, maximum and median (see
Figure~\ref{fig:5} or Figure~\ref{fig:180619} above.)
This design leads
to efficient bare median (19 links) and
full sorting (25 links)
on nine-element arrays;
see Figure~\ref{fig:8} below.

\begin{figure}[H]
\begin{center}
\begin{tabular}{c c c}
\includegraphics[scale=0.110]{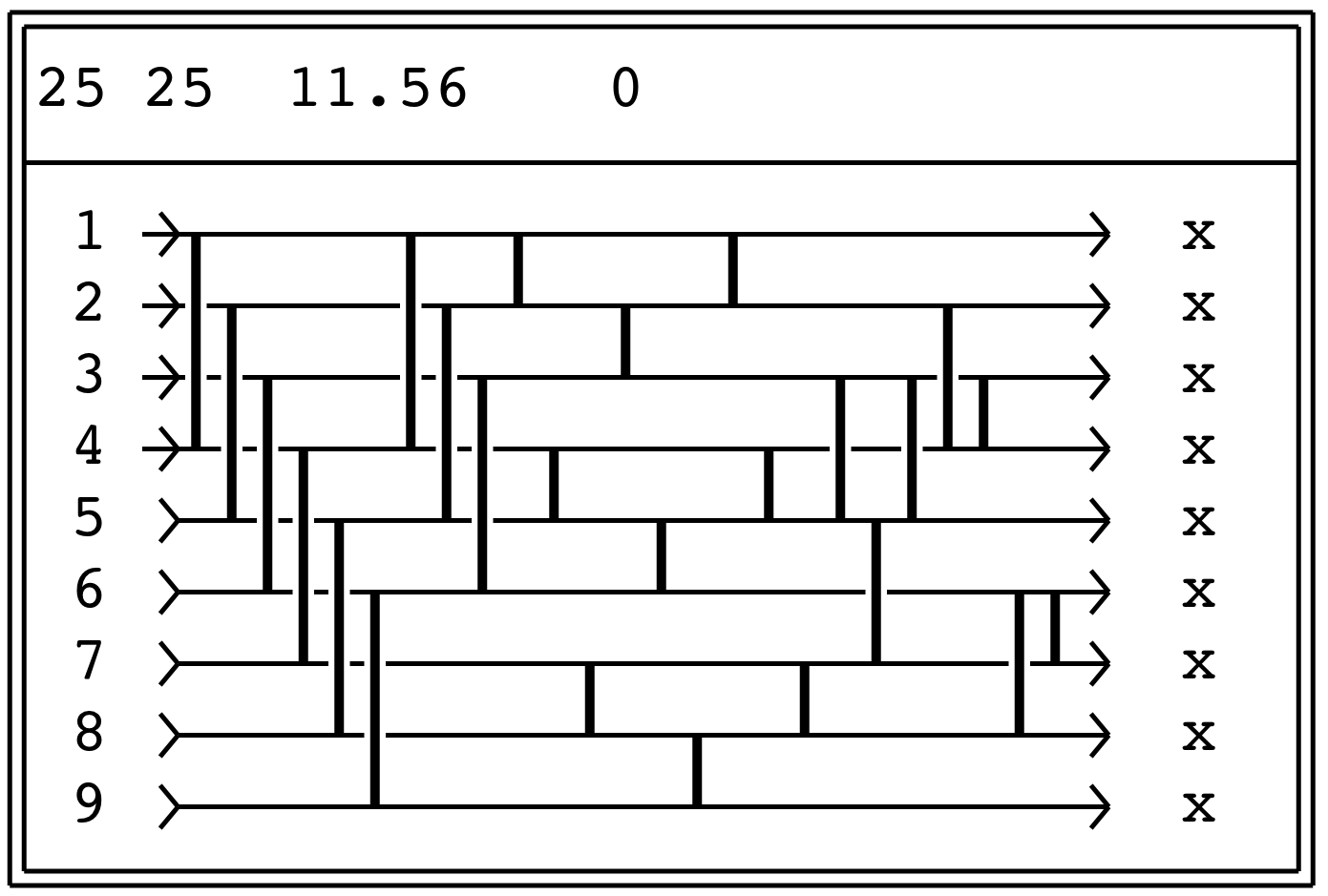}
& &
\includegraphics[scale=0.115]{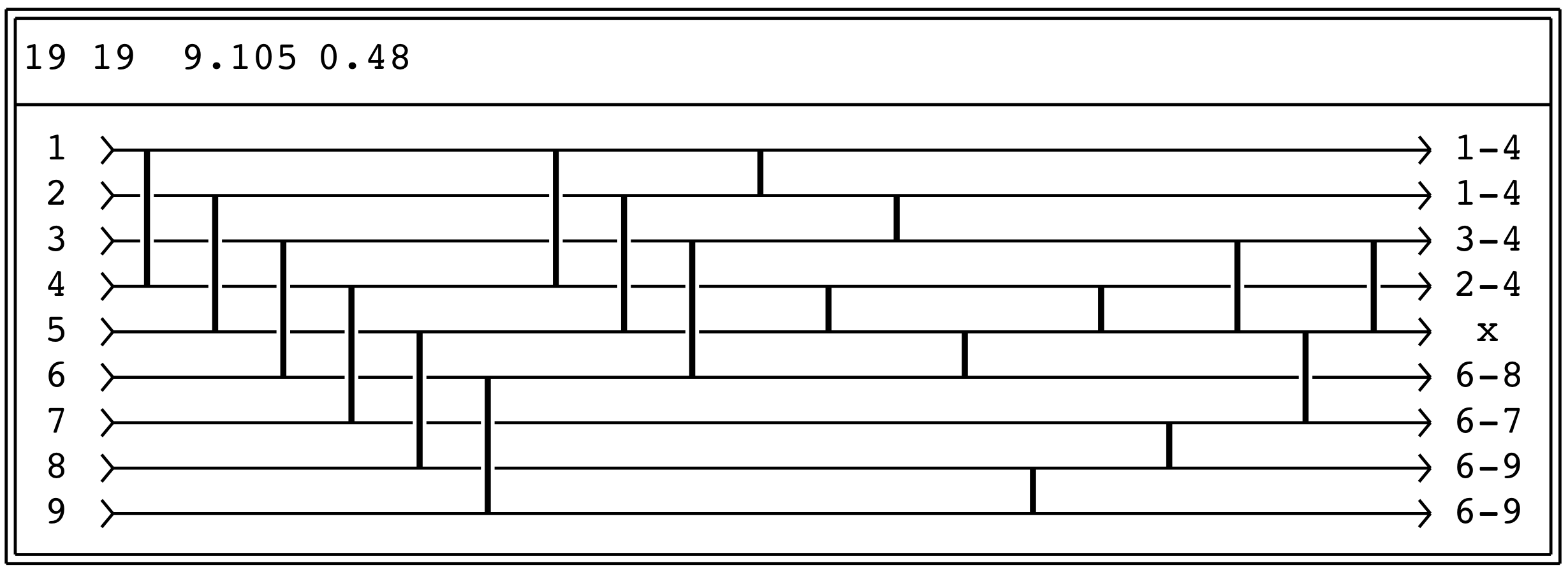}    \\
(a) & & (b)
\end{tabular}
\end{center}
 \caption{\small
 Two sorting networks derived from that in
 Figure~\ref{fig:180619}.
 (a) corresponds to the fully sorted version by adding the links
 $(23),(34),(68),(67)$ at the end.
 (b) corresponds to the bare median version by deleting the first
 $(1,2)$ link and the last $(7,8)$ link.
  }
\label{fig:8}
\end{figure}

Exchanges on 21 links occur when presented reverse sorted input,
and (see Figure~\ref{fig:5}) $9\sfrac{27}{35}\approx 9.771$ times
on average which gives  a relative frequency of
$(9\sfrac{27}{35})/21 = \sfrac{114}{245} \approx 47\%$.

\subsection{New design using 3-ops}

Substitution of $\sthree$ by its reworked 3-op version gives
the network in Figure~\ref{fig:7} below.

\begin{figure}[H]
\begin{center}
\includegraphics[width=0.6\textwidth]{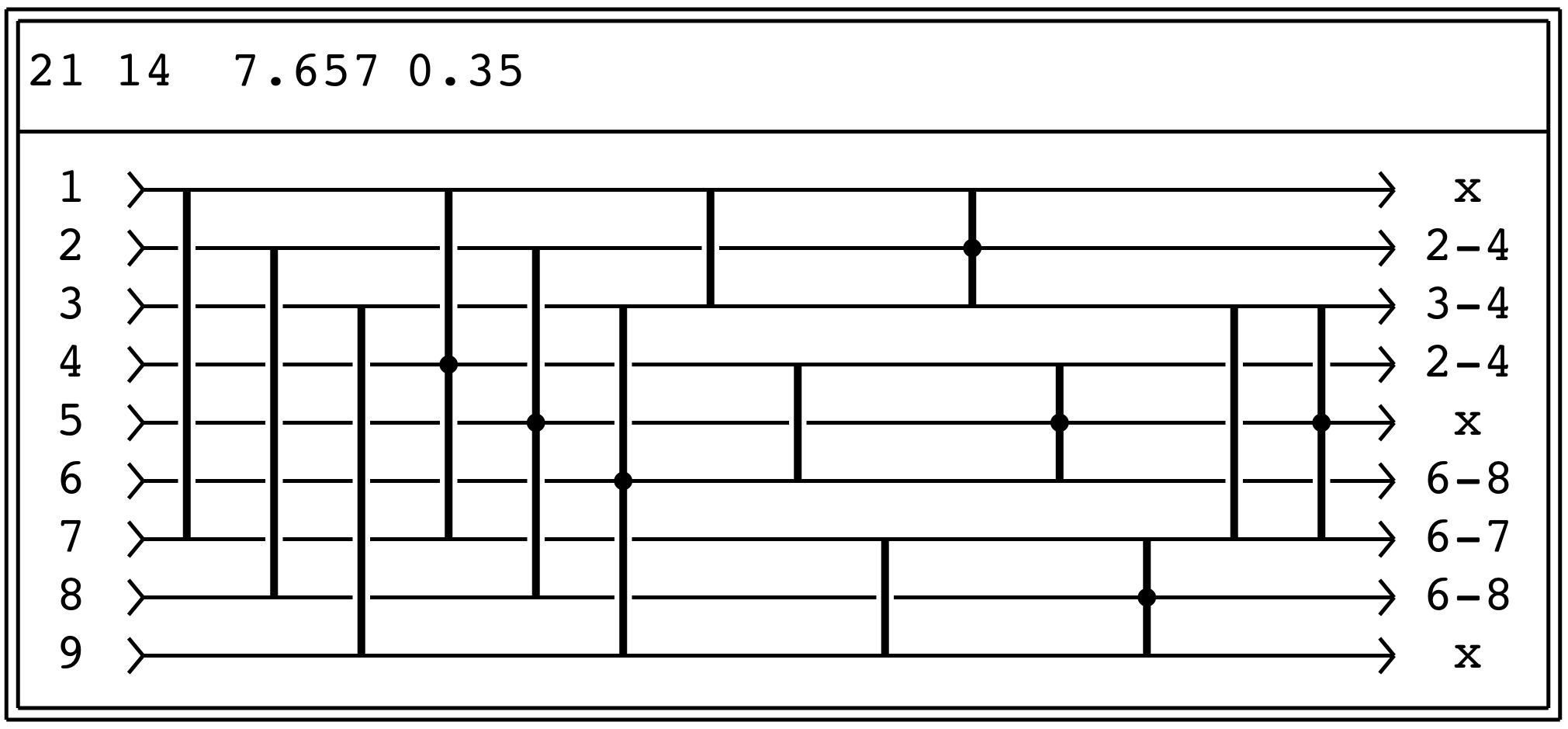}
\end{center}
 \caption{\small
 A new network based on 3-ops, featuring 6 stages.
}
\label{fig:7}
\end{figure}

Full sorting in 25 links\footnote{Note that $25$ is also the
minimum number required to sort a 9-network; see \cite[Table~(11)
on page~226 in Section~5.3.4]{Knuth}.} adds the four CE elements
$(2,4)(3,4)(6,8)(6,7)$;
Bare median finding in 19 links omits
$(1,2)$ and $(8,9)$ from the 2-ops $(1,2,3)$ and $(7,8,9)$,
respectively.
Note that these omissions cannot be made directly on the original net
without permutations to allow for median. The final networks are
presented in Figure~\ref{fig:8+} below.

\begin{figure}[H]
\begin{center}
\begin{tabular}{c c c}
\includegraphics[scale=0.097]{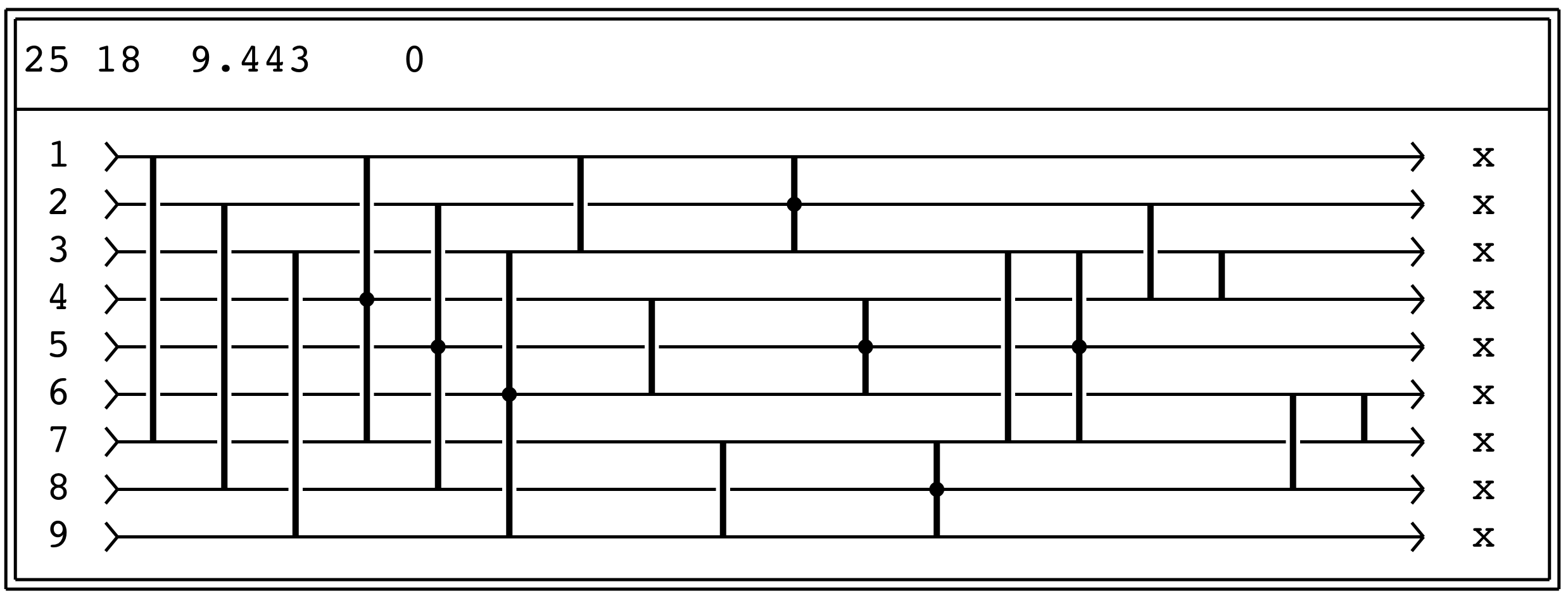}
& &
\includegraphics[scale=0.097]{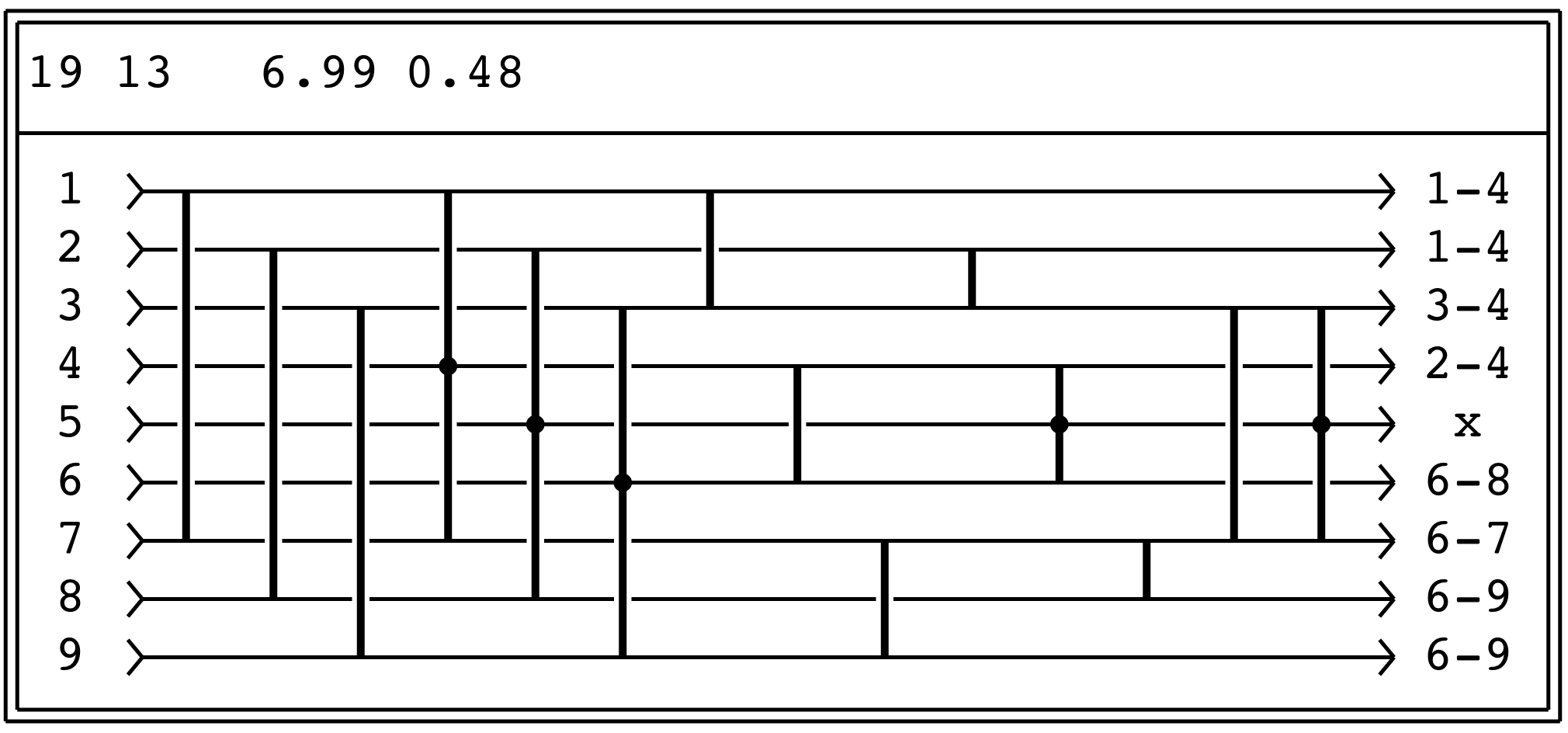}    \\
(a) & & (b)
\end{tabular}
\end{center}
 \caption{\small
 Two sorting networks derived from that in Figure~\ref{fig:7}.
 (a) corresponds to the fully sorted version by adding the links
 $(23),(34),(68),(67)$.
 (b) corresponds to the bare median version by deleting the links
 $(1,2)$ and $(8,9)$ from the 2-ops
 $(1,2,3)$ and $(7,8,9)$ respectively.
  }
\label{fig:8+}
\end{figure}

\subsection{Comparison}

We summarize the benefits of the new design versus the old design
in Table~\ref{t:0}:

\begin{center}
\begin{tabular}[t]{@{}rrrrcrrrcrrr@{}}
\toprule
Design &\multicolumn{3}{c}{Median only} & \phantom{abc}
&\multicolumn{3}{c}{Min/Med/Max} & \phantom{abc}
&\multicolumn{3}{c}{Fully sorted}\\
\cmidrule{2-4} \cmidrule{6-8} \cmidrule{10-12}
&\multicolumn{2}{c}{Swaps} &\multicolumn{1}{c}{Stages}
&\phantom{abc}
&\multicolumn{2}{c}{Swaps} &\multicolumn{1}{c}{Stages}
&\phantom{abc}
&\multicolumn{2}{c}{Swaps} &\multicolumn{1}{c}{Stages}\\
\cmidrule{2-3} \cmidrule{6-7} \cmidrule{10-11}
& Avg & Max & & & Avg & Max & & & Avg & Max\\
old & 9.105 & 19 &9 & & 9.771 & 21 & 9 & & 11.56 & 25 & 11\\
new & 6.99 & 13 & 6 & & 7.657 & 14 & 6 & & 9.443 & 18 & 8\\
\midrule
new/old & 76.8\% & 68.4\% &  66.7\% & & 78.4\% & 66.7\% & 66.7\%
& & 81.7\% & 72.0\% &  72.7\%\\
\bottomrule
\end{tabular}
\captionof{table}{Summarized statistics of the old and the new design for
sorting $9$-element networks.}
\label{t:0}
\end{center}

We conclude our discussion by listing Schwiebert's sorting
network which is known to have the minimum number of stages/delay
time. Note that the average and maximum
number of comparisons is considerably higher.

\begin{figure}[H]
\begin{center}
\includegraphics[width=0.5\textwidth]{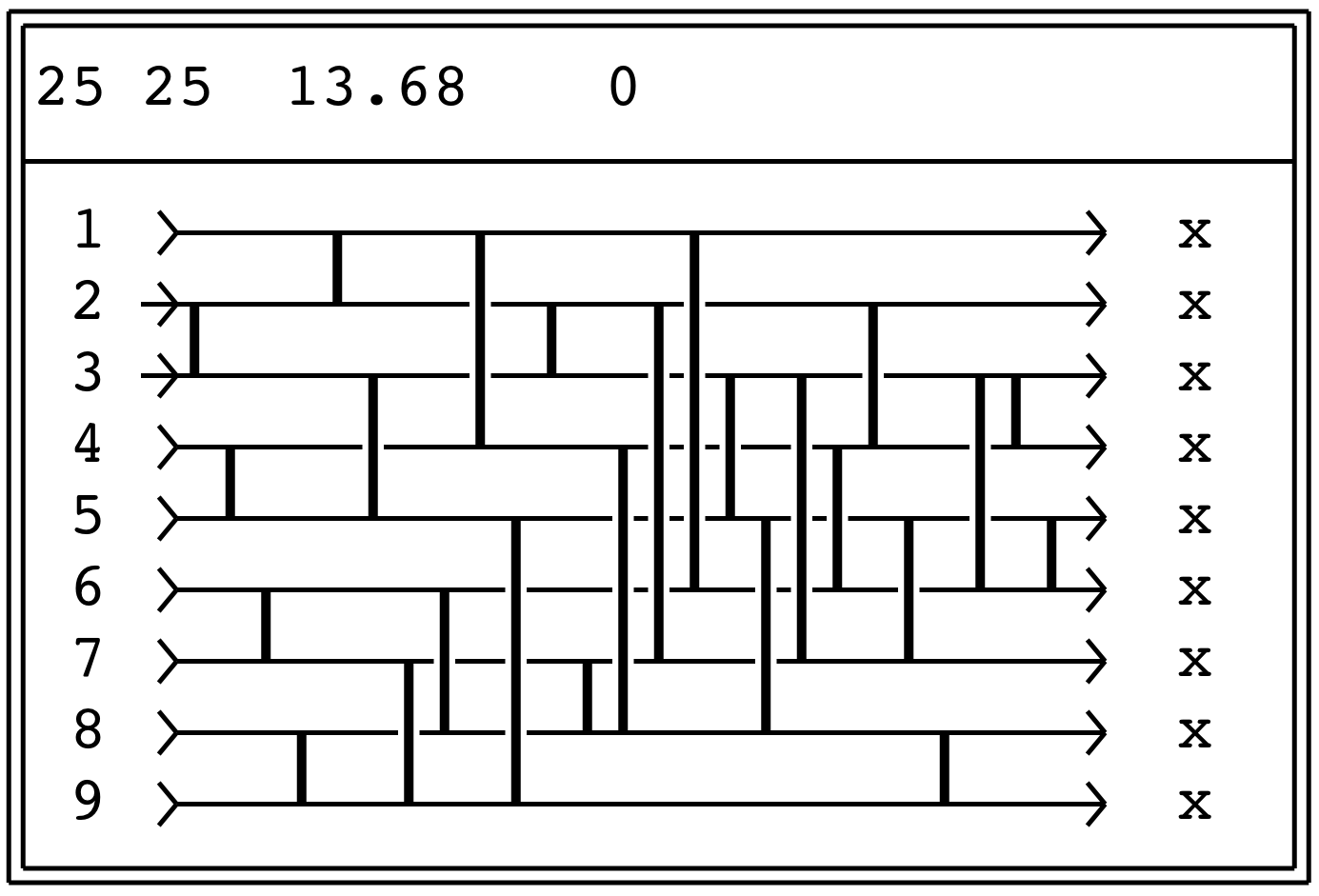}
\end{center}
 \caption{\small
Loren Schwiebert's 25-link sorting network for 9 elements taken from
\cite[Figure~51 for $n=9$]{Knuth} with delay time $7$, i.e., $7$
stages.
}
\label{fig:knuth51nis9}
\end{figure}

\section{Intuition and Extension}

To some the lowered statistics seem surprising. Typical reasoning follows: "to
sort three, order any two, the third draws a bye. Bye enters and plays to
establish an overall winner (WLOG, loser); third step ranks the runner's up."
While accurate when elements are chosen freely, as a memory-based method each
element carries a rank implicit in its index. CE sorting ranks all elements,
chosen pairs are not isomorphic. In the case of $N=3$ we may disambiguate thus:
``Choosing any two at most omits the median, our set is are guaranteed to hold
at least the min or max. Choosing and sorting the outermost elements therefore
guarantees at least one of max or min landing in its final position. A single
swap will place the remaining extrema; both extreme must be treated (a 2-op).
With both extrema placed, median is properly placed by process of
elimination''.

Students of computer science will recognize Figure~\ref{fig:1}(a)
as bubble-sort on three elements
where the initial link leaves the maximum element in central position
$2/3$ of the time (for the remaining $1/3$, it started in proper
position and will not move), thereby encumbering subsequent links
the task of floating it into its outer position.
By comparison, the non-interfering method is seen as a selection or insertion
sort whose improved swapping statistics are well-known. It is ironic such
study of CE networks is rare and that the total number of links (or stages) has
remained the sole performance metric for so long.

Conceptually, the longer prefacing element link in Figure~2(a)
serves as a guard in assuring that no double-exchange occurs,
allowing fusion into one 2-op.  The final links might not be unit
length, as seen in the improved sort/median on nine elements.
The guard needs only protect (operate on) the outer two unshared elements.
As a second and more interesting means of extension, we envision a
3-op of higher order, formed by allowing two adjacent 2-ops to
execute concurrently. We explore the latter now.

\section{Sorting $N=4$}

\begin{figure}[H]
\begin{center}
\begin{tabular}{c c c}
\includegraphics[scale=0.15]{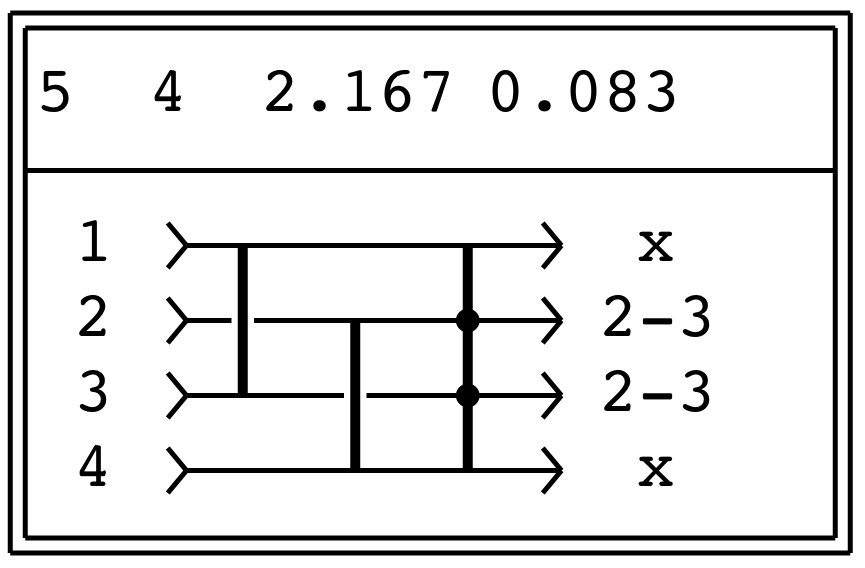}
& & \includegraphics[scale=0.15]{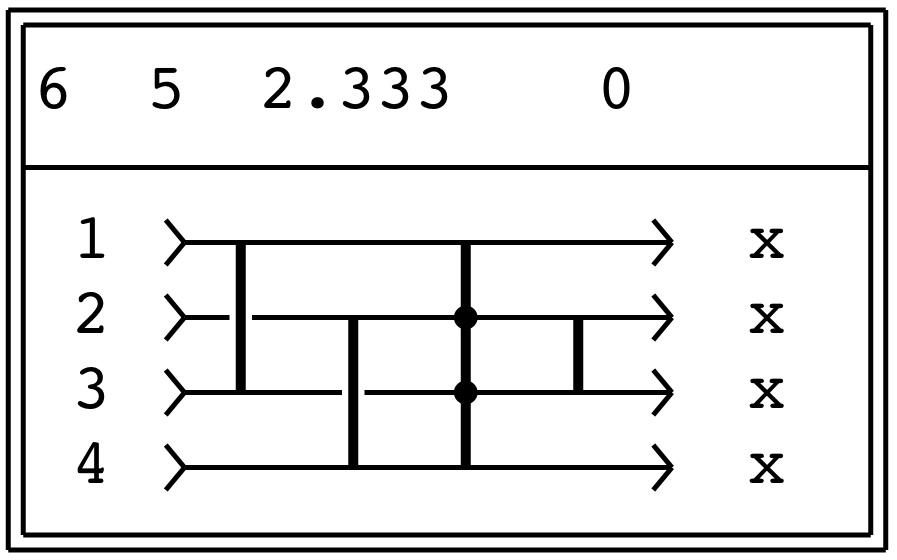}
\\
(a) & & (b)
\end{tabular}
\end{center}
 \caption{\small
 A sorting network for $N=4$ identifying min/max in (a) on the left,
 and fully sorting in (b) on the right.
  }
\label{fig:10}
\end{figure}

A trial network (see Figure~\ref{fig:10}(a))
overlaps two optimal 3-sorts, sharing a central element,
whose sorting statistics and testing reveal a maximum
swap of four elements,
occurring only with the center link $(2,3)$ of the 3-op $(1,2,3,4)$,
which consists of 3 links without conflict:
$(1,2)$, $(2,3)$, and $(3,4)$
where $(2,3)$ is either non-active or the \emph{only} active link.
Unfortunately, this
network fails to fully sort in five steps:
$1/6$ of the time, the central elements
are transposed, requiring a sixth link as seen in Figure~\ref{fig:10}(b).
This is both undesirable and surprising:
the final link seemingly replicates the work
concluded immediately its left. There are two remedies to this. Both rely on
the fact that the central link in the 3-op might not execute, yet its operation
is essential to attain a full sort.

\begin{figure}[H]
\begin{center}
\begin{tabular}{c c c}
\includegraphics[scale=0.15]{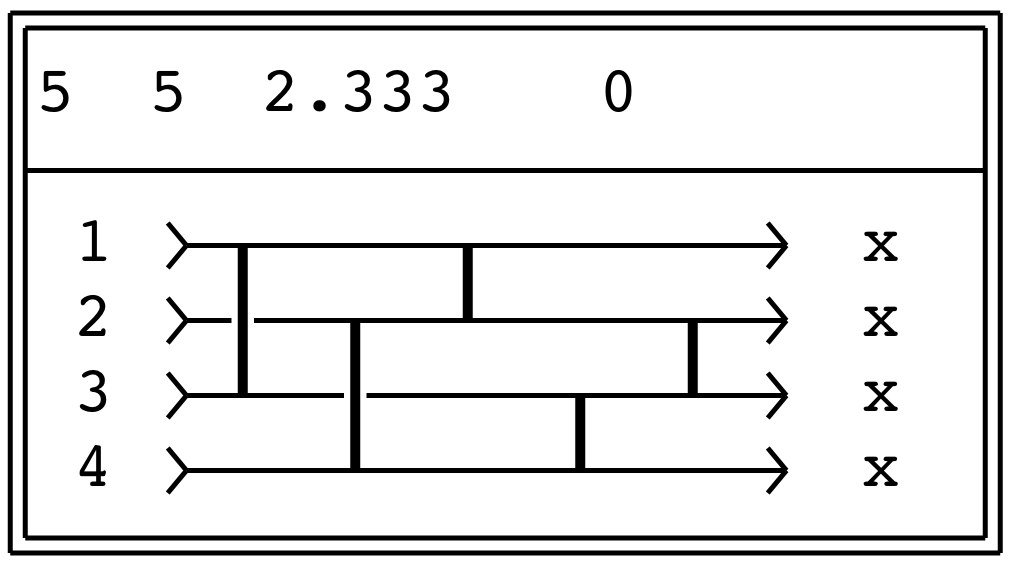}
& & \includegraphics[scale=0.15]{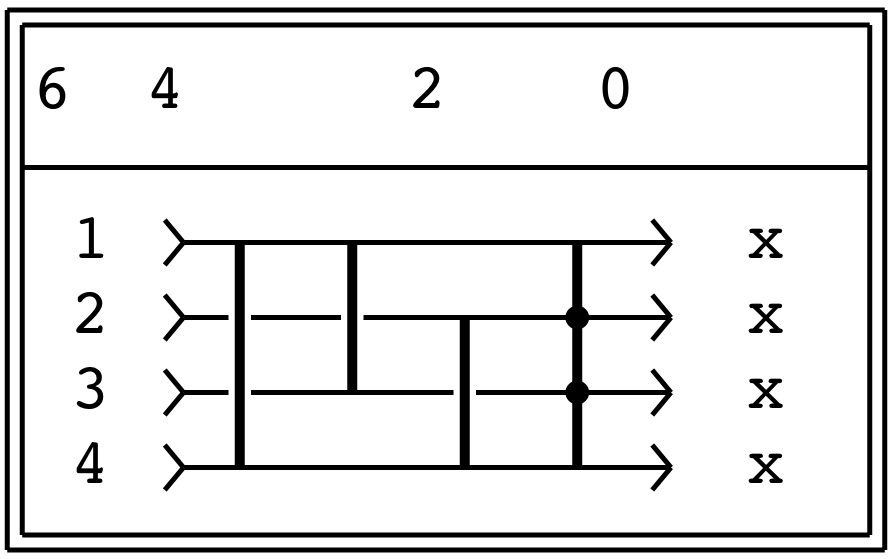}
\\
(a) & & (b)
\end{tabular}
\end{center}
 \caption{\small
(a): Swap frequencies are 
$\sfrac{1}{2}$ at links 1,2,3,4;
$\sfrac{1}{3}$ at link 5.
(b): Swap frequencies are 
$\sfrac{1}{2}$ at link~1;
$\sfrac{1}{3}$ at links~2 and 3;
$\sfrac{1}{4},\sfrac{1}{2},\sfrac{1}{4}$ at
$(1,2),(2,3),(3,4)$ from the 3-op $(1,2,3,4)$, respectively.
  }
\label{fig:11}
\end{figure}

Our first solution breaks the 3-op into its constituent links. Since
they do not interfere, we are afforded any ordering of its components.
We first choose to place the link to its right, giving rise to
redundant right-most central links (not shown). We retain one only,
giving the familiar solution Figure~\ref{fig:11}(a), sans N-ops.
The second solution, depicted in Figure~\ref{fig:11}(b),
recognizes that the additional link is necessary in the case the
3-op is fully active (i.e., two outer swaps),
its actions then juxtaposing two (as
yet unordered) elements against each other into adjacent central position.
As before,
we guard against interference by employing a prefacing link whose two
endpoints are the outer endpoints (elements) of the two links it
protects.

The second solution (in Figure~\ref{fig:11}(b))
is in fact the $\mathcal{O}(n^2)$
pair-wise comparison of all elements,
taken in descending order, which necessarily ranks all elements.
While not a
viable solution for large $N$, for $N=4$ it
sits close to the theoretical minimum
(5 links, see Figure~\ref{fig:12} below)
and rewards the implementor with surprisingly
good statistics for
average swaps (2) while nonetheless executing in at most three hardware stages.

Large networks whose size $N$ is a power of two, have many prefacing
links whose sorting statistics show a $50/50$ probability of exchange.
The transposition operator may be applied to any such link (and to
any implicated links lying to its left) without changes to the
average sorting statistics. The topological change may instead
succeed in reducing the maximum number of exchanges.
Applied to the network seen in Figure~\ref{fig:11}(a), we obtain
the network depicted in Figure~\ref{fig:12} below,
where the maximum number of exchanges dropped from 5 to 4.

\begin{figure}[H]
\begin{center}
\includegraphics[width=0.3\textwidth]{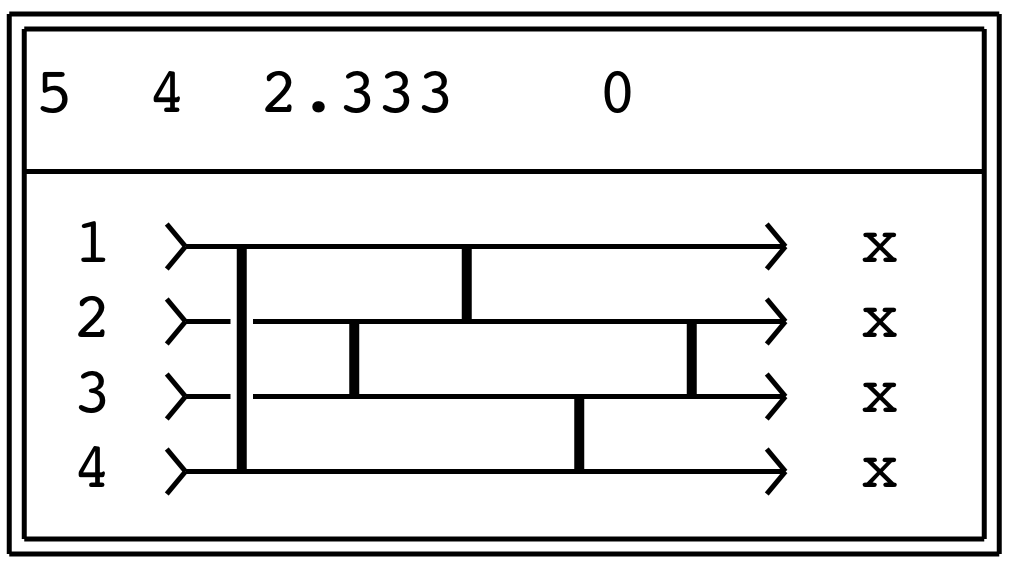}
\end{center}
 \caption{\small
A 5-link sorting network for 4 elements.
The last link has swap probability of $\sfrac{1}{3}$;
all other links have swap probability of $\sfrac{1}{2}$.
}
\label{fig:12}
\end{figure}

The worst-case
input occurs when reordering $(3,4,1,2)$
and in this case all save but the
final link are active.
A breakdown of total swap counts appears below:

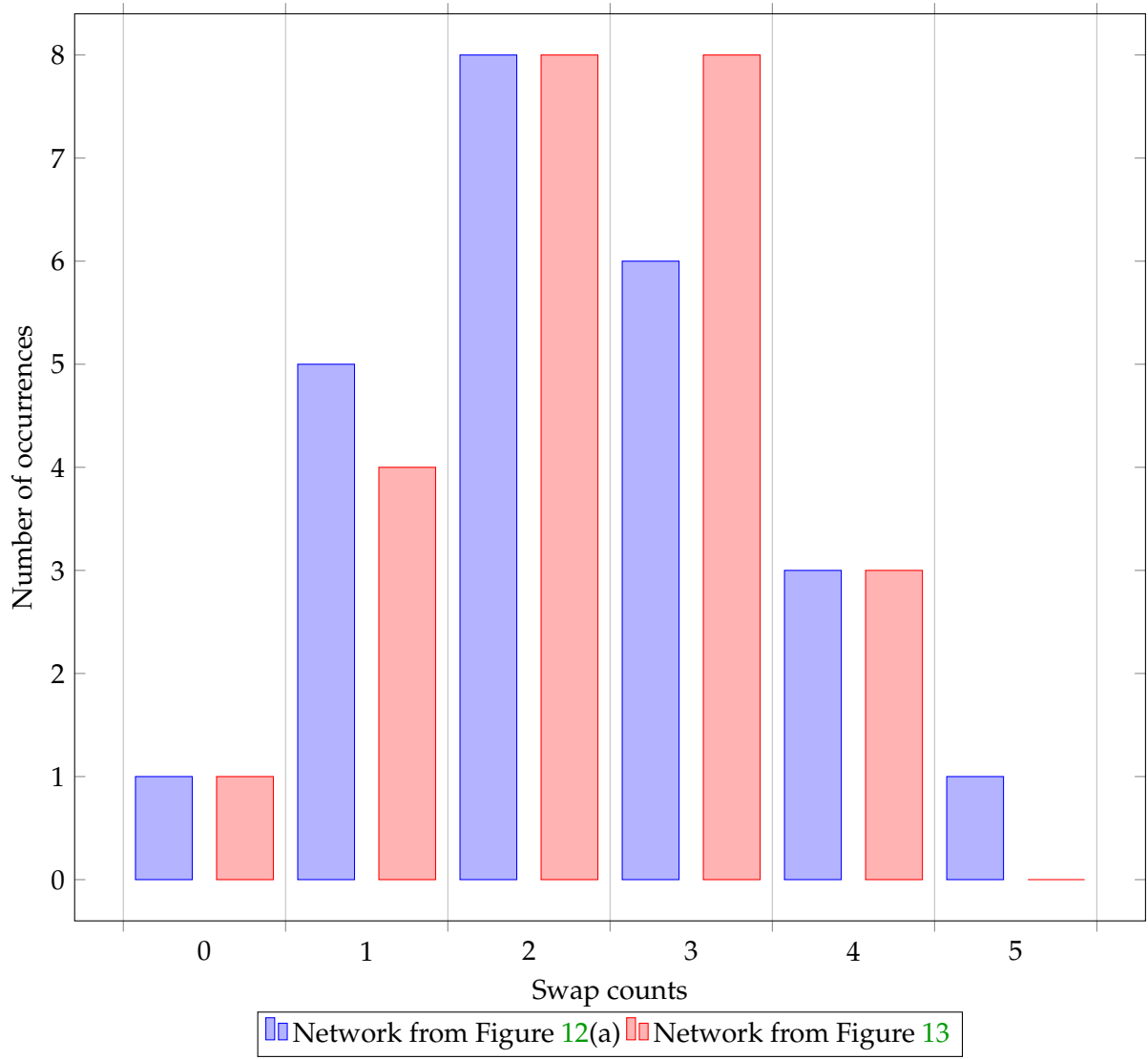
\begin{figure}[H]
\begin{tikzpicture}
\begin{axis}[
	x tick label style={
		/pgf/number format/1000 sep=},
  xlabel=Swap counts,
	ylabel=Number of occurrences,
	enlargelimits=0.05,
	legend style={at={(0.5,-0.1)},
	anchor=north,legend columns=-1},
	ybar interval=0.7,
]
\addplot
	coordinates {(0,1) (1,5) (2,8) (3,6) (4,3) (5,1) (6,0)};
\addplot
	coordinates {(0,1) (1,4) (2,8) (3,8) (4,3) (5,0) (6,0)};
\legend{Network from Figure~\ref{fig:11}(a),Network from Figure~\ref{fig:12}}
\end{axis}
\end{tikzpicture}
 \caption{\small
 A histogram for representing the number of occurrences with respect to
number of exchanges/swaps
 required for all possible
 inputs for the two networks from
 Figures~\ref{fig:11}(a) and Figure~\ref{fig:12}.
 The input requiring 5 exchanges for the network from Figure~\ref{fig:11}(a)
is $(4,2,3,1)$.}
\end{figure}

Note that the reappearance of the central link $(2,3)$ in
Figure~\ref{fig:12} implies that a "needless" i.e. redundant
comparison might take place, as when presented the input $(1,3,2,4)$
requiring merely one central transposition. This cannot be helped;
the network is statistically optimal. In the extreme case, sorted
data requiring no exchanges will nonetheless compare at all link
positions, including "redundant" links -- such is the topology of
CE networks. Attempts to mitigate this, such as (software) early
termination as when reaching the threshold of maximum exchange are
not cost effective: they must both count and test in order to detect
this condition at a cost often greater than a few additional link
comparisons.

The network depicted in Figure~\ref{fig:12}
also sheds light into the subtleties of conditional probability.
While there are $4!= 24$ possible inputs permutations, the first four links
cannot each successively half the size of the solution space:
$24$ is not divisible by $2^4=16$.
While links 3 and 4 both execute independently with $50/50$ exchange
statistics, they are nonetheless conditionally bound.
If link 3 does execute,
then there is a $\sfrac{2}{3}$ chance that link 4 does not,
and vice versa.
In effect, when "double winner" (max of all four) is found
by exchanging element 2 with 3, previous losers to this element may
be ascribed less demotion; they are more apt to retain a superior
position and not trigger a link 4 exchange. Intuition aside, careful
analysis of exchange statistics bears this all out.

Finally, we compare the network in Figure~\ref{fig:12} to the
following one found in \cite[Figure~44]{Knuth}:
\begin{figure}[H]
\begin{center}
\includegraphics[width=0.3\textwidth]{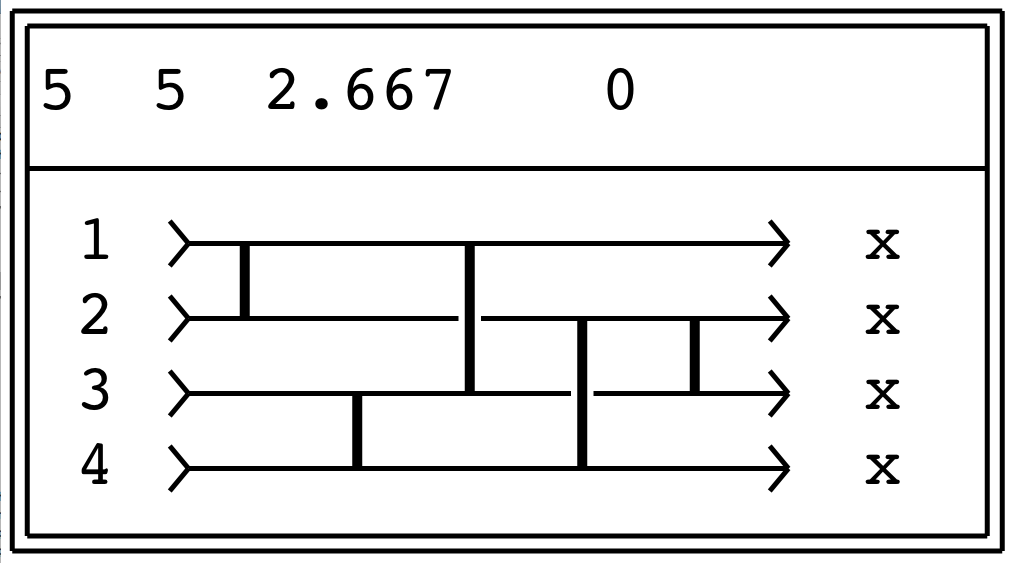}
\end{center}
 \caption{\small
A 5-link sorting network for 4 elements taken from
\cite[Figure~44]{Knuth}.
The last link has swap probability of $\sfrac{2}{3}$;
all other links have swap probability of $\sfrac{1}{2}$.
}
\label{fig:knuth44}
\end{figure}
Note that this network does not compare as well as the one in
Figure~\ref{fig:12}; it requires 5 swaps and the average number
of swaps is higher as well.

\section{Software Implementation}

Reduction in total exchanges without the use of conditional directives is the
greatest benefit of CONEX techniques. In the case of 2-ops and 3-ops a simple
software addition can reduce the explicit cost of an unnecessary comparison:

\begin{verbatim}
  #define sort2(a,b)   if ((a) < (b)) { (a) ^= (b); (b) ^= (a); (a) ^= (b) }
  #define op2(a,b,c)   if (sort2(a,b)) else sort2(b,c)
  #define sort3(a,b,c) sort2(a,c); op2(a,b,c)
\end{verbatim}

in which the "else" elides a non-interfering CE element. The form shown above
provides exchanges without resort to additional variables / registers. If we
allow their inclusion the difference implicit in the first comparison can be
retained to speed the body of the exchange,
as shown elsewhere \cite{Paeth90}.
\begin{verbatim}

  #define sort2(a,b)   if ((t = (a) - (b)) < 0) { (b) += t; (a) -= t); }
\end{verbatim}

With 3-ops, an exchange of the central element will elide possible swaps by
the outlying ``wings''
\begin{verbatim}

  #define op3(a,b,c,d)   if (sort2(b,c)) else { sort2(a,b); sort2(c,d); }
\end{verbatim}
These macros expand as in-line code, removing subroutine overhead and form the
basic primitives used in software implementations of NICE networks.

\section{Higher Order Networks}

For $N$ odd,
we explore methods of median finding,
for $N$ even, we consider fully
sorting networks.

\subsection{Sorting $N=5$}

An efficient method for $N=5$ is elusive:
network sizes which are a power of two
show a high degree of symmetry,
sizes $2^n+1$ must accommodate the new addition.
Theory prescribes 7 comparisons (as $5! = 120 < 128 = 2^7$),
and is realizable
in careful practice. For CE implementations, a worst case figure of 10 links is
an upper bound, this is the \emph{triangle number} $T(4)=4(4+1)/2$
enumerating all pairwise
comparisons of five elements.
In practice, nine links is the minimum CE
implementation. Exhaustive searching reveals that a simple elimination of one
maximal element (four steps) followed by a sort4 (five links) gives the best
sorting statistics:
\begin{figure}[H]
\begin{center}
\includegraphics[width=0.5\textwidth]{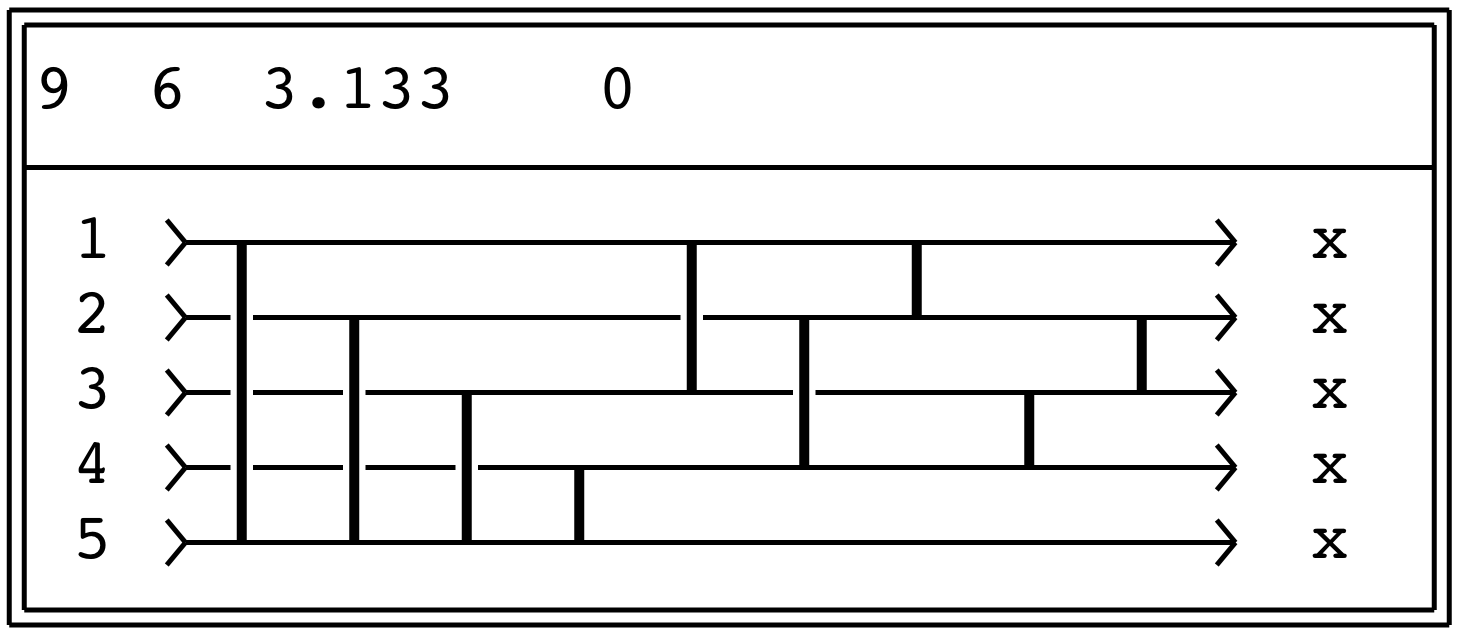}
\end{center}
 \caption{\small  
 A 9-link network to sort 5 elements.
The probabilities of swapping are:
$\sfrac{1}{2}$ at link~1,
$\sfrac{1}{3}$ at link~2,
$\sfrac{1}{4}$ at link~3,
$\sfrac{1}{5}$ at link~4,
$\sfrac{1}{3}$ at link~5,
$\sfrac{5}{12}$ at link~6,
$\sfrac{2}{5}$ at link~7,
$\sfrac{2}{5}$ at link~8,
$\sfrac{3}{10}$ at link~9.
This results in 
$3\sfrac{2}{15} = 3.1\overline{3}$
swaps on average.
}
\label{fig:14}
\end{figure}

Note that the final sort is of the style of that one in
Figure~\ref{fig:11}(a),
and not the one in Figure~\ref{fig:12}.
While not strictly a NICE
network, a compact software implementation for
$N=5$ is a good project for future work.

\subsection{Sorting $N=6$}

For $N=6$, three-sort the elements $(1,3,5)$ and
symmetrically on $(2,4,6)$, after sorting $(1,5)$ and $(2,6)$.
Note that in each case a central 2-op occurs on element 3 (4).
We now center these
two elements by again using a 2-op,
but this time against the outer interval
formed by the opposing 3-sort;
that is, the 2-ops $(1,4,5)$ and $(2,3,6)$.
These leads to a highly parallel execution model
in which the central two elements
are located at $(3,4)$ albeit unordered.
By the ranking nature of CE networks,
elements $(1,2)$ and $(5,6)$ are also properly partitioned,
albeit unsorted.
Three concurrent links at $(1,2)$, $(3,4)$ and $(5,6)$
thereby conclude the sort (see Figure~\ref{fig:15}):

\begin{figure}[H]
\begin{center}
\includegraphics[width=0.5\textwidth]{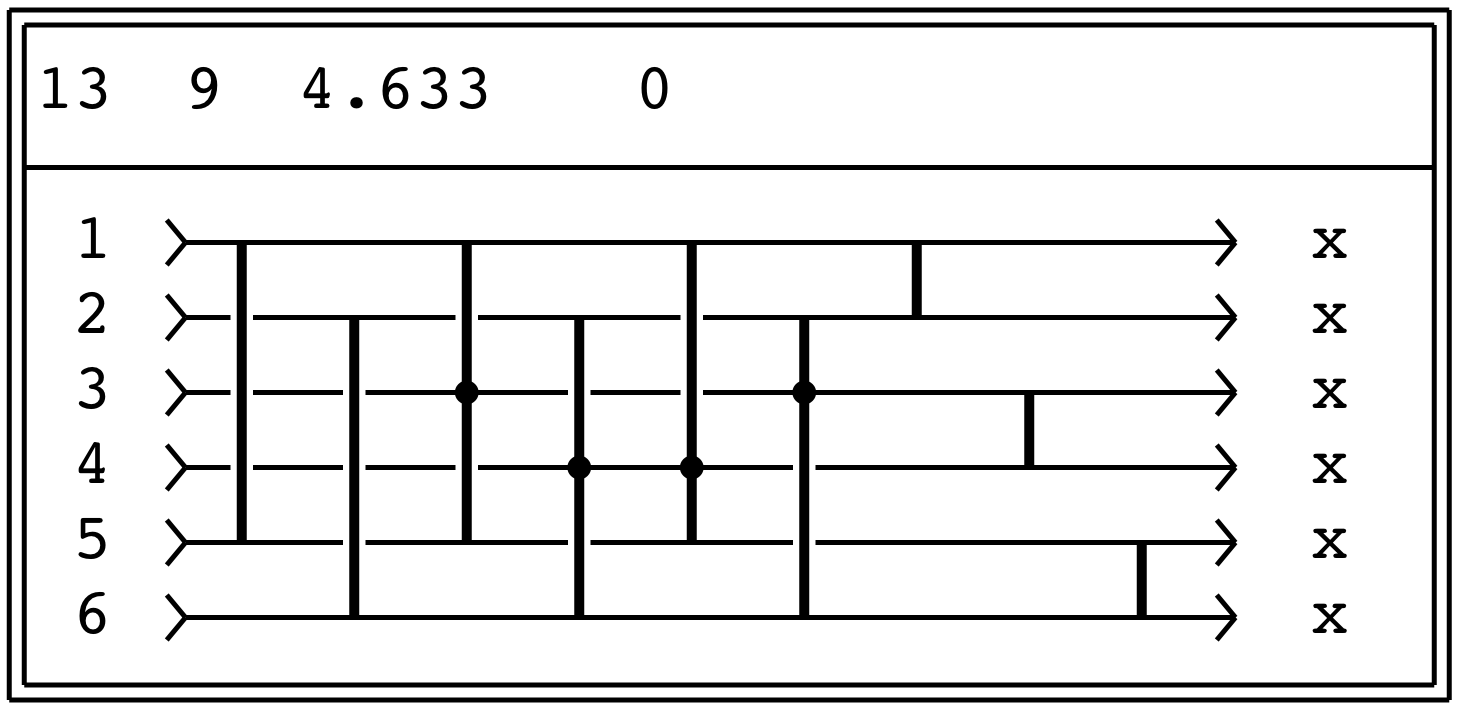}
\end{center}
 \caption{\small
 A network to sort $6$ elements.
}
\label{fig:15}
\end{figure}

\begin{figure}[H]
\begin{center}
\includegraphics[width=0.5\textwidth]{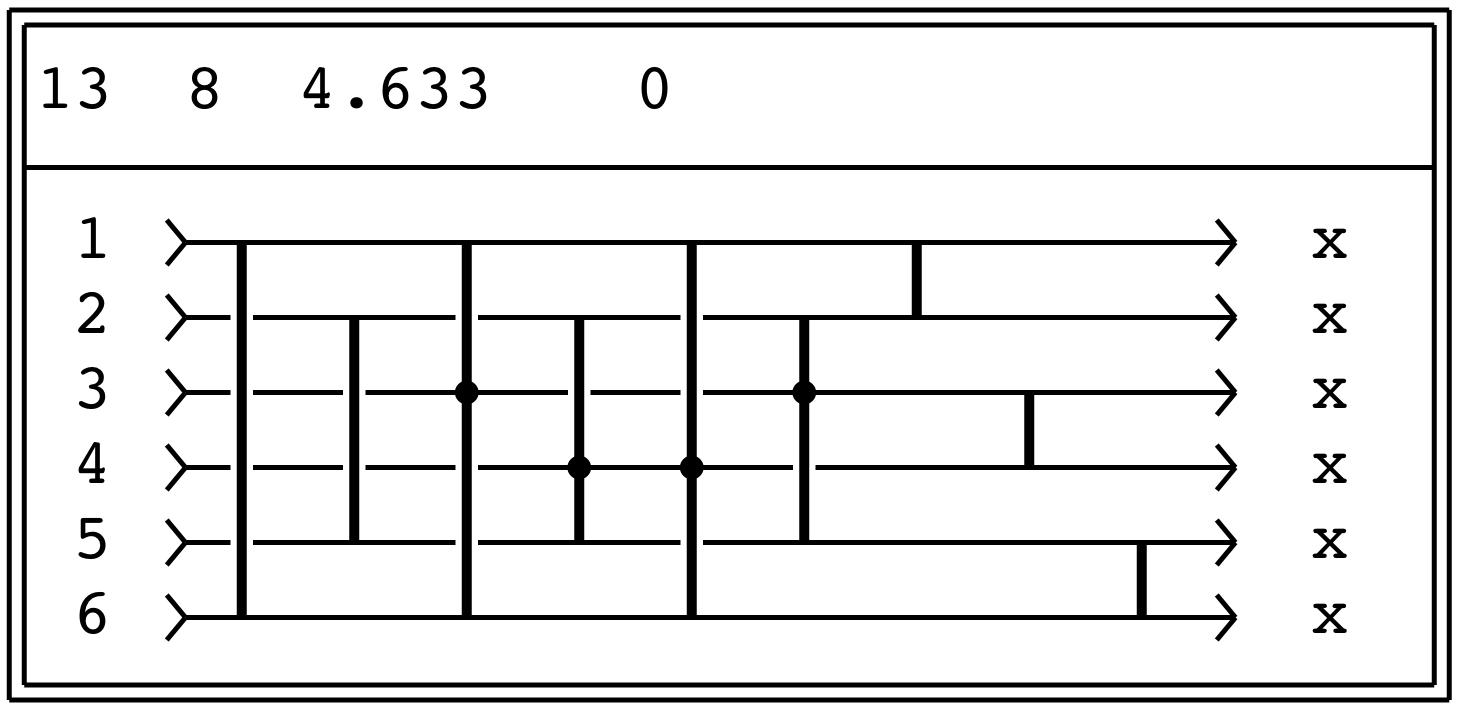}
\end{center}
 \caption{\small
 A rework of the network from Figure~\ref{fig:15} to sort $6$ elements
 with improved worst-case performance.
}
\label{fig:16}
\end{figure}

A rework of the network from Figure~\ref{fig:15}
presented in Figure~\ref{fig:16}
preserves the overall exchange statistics while lowering
the worst-case performance to 8 swaps.

Note that both networks complete in only
four stages with the two central stages involving all elements concurrently.

If the goal is to reduce average swaps in reduced links,
the network in Figure~\ref{fig:17} will do that,
at the expense of additional stages:

\begin{figure}[H]
\begin{center}
\includegraphics[width=0.5\textwidth]{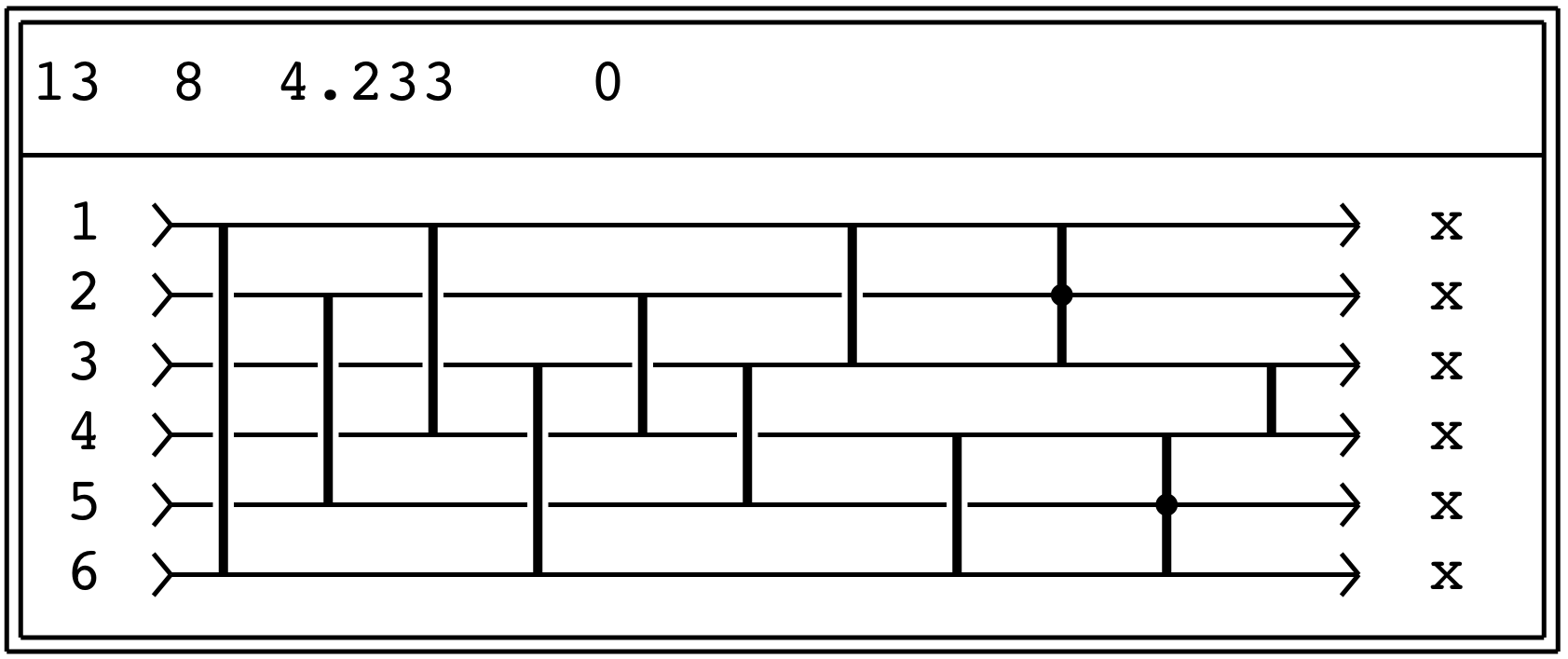}
\end{center}
 \caption{\small
 A network to sort $6$ elements with minimum average-swap performance.
 }
\label{fig:17}
\end{figure}

Applying basic techniques
(symmetric min/max, applied successively)
leads to the network depicted in Figure~\ref{fig:18} which
sorts six elements in just 12 links -- the minimum:

\begin{figure}[H]
\begin{center}
\includegraphics[width=0.5\textwidth]{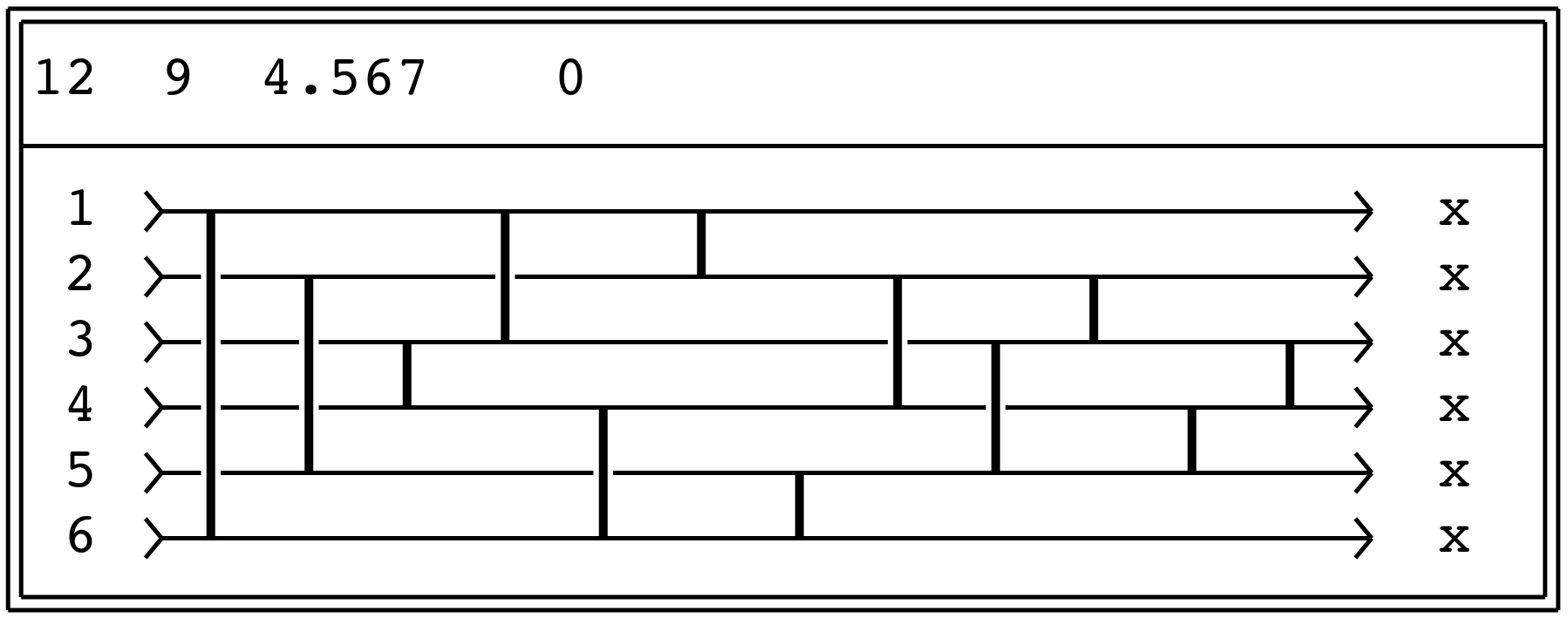}
\end{center}
 \caption{\small
 A network to sort $6$ elements with the minimum number (12) of links.
 }
\label{fig:18}
\end{figure}
The user may achieve the smallest total number of swaps (8 compared to 9)
at the
cost of a slightly increased average number of swaps.
(This trade-off with mild
change in topology symmetry is seen with large networks in general.)

\begin{figure}[H]
\begin{center}
\includegraphics[width=0.5\textwidth]{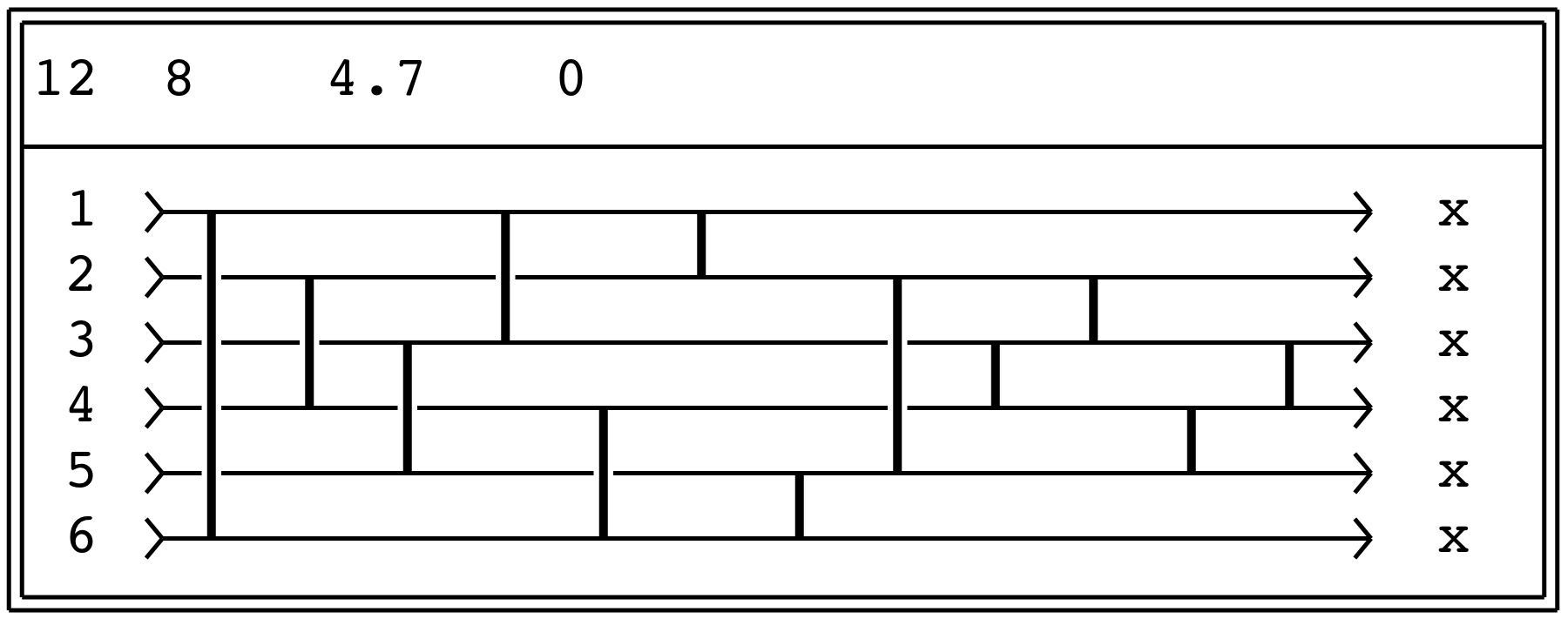}
\end{center}
 \caption{\small
 Another network to sort $6$ elements with the minimum number (12) of links
 but reduced worst-case scenario at the cost of slightly worse average
 swap performance.
 }
\label{fig:19}
\end{figure}

Finally, let us now present
Gerald Norris Shapiro's network (taken from \cite[Figure~51]{Knuth})
for sorting six elements in Figure~\ref{fig:20}:

\begin{figure}[H]
\begin{center}
\includegraphics[width=0.5\textwidth]{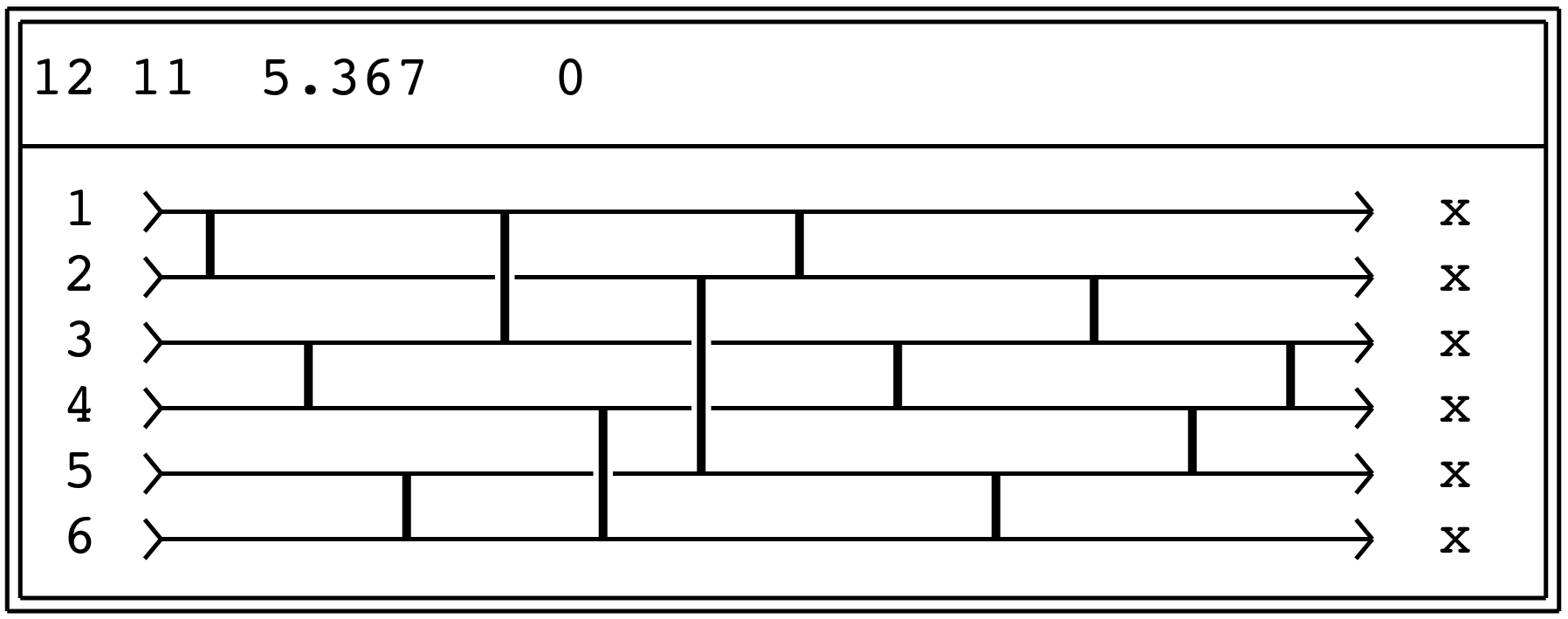}
\end{center}
 \caption{\small
 Shapiro's network for sorting six elements, requiring 5 stages and 12 links.
 }
\label{fig:20}
\end{figure}

Given the minimum number of links (12), Shapiro's network has the
minimum number of stages (5); however, in terms of worst case scenario
and average number of swaps, the networks in Figure~\ref{fig:18} and
Figure~\ref{fig:19} do better at the expense of just one more stage.

\subsection{Sorting $N=7$}

\begin{figure}[H]
\begin{center}
\includegraphics[width=0.6\textwidth]{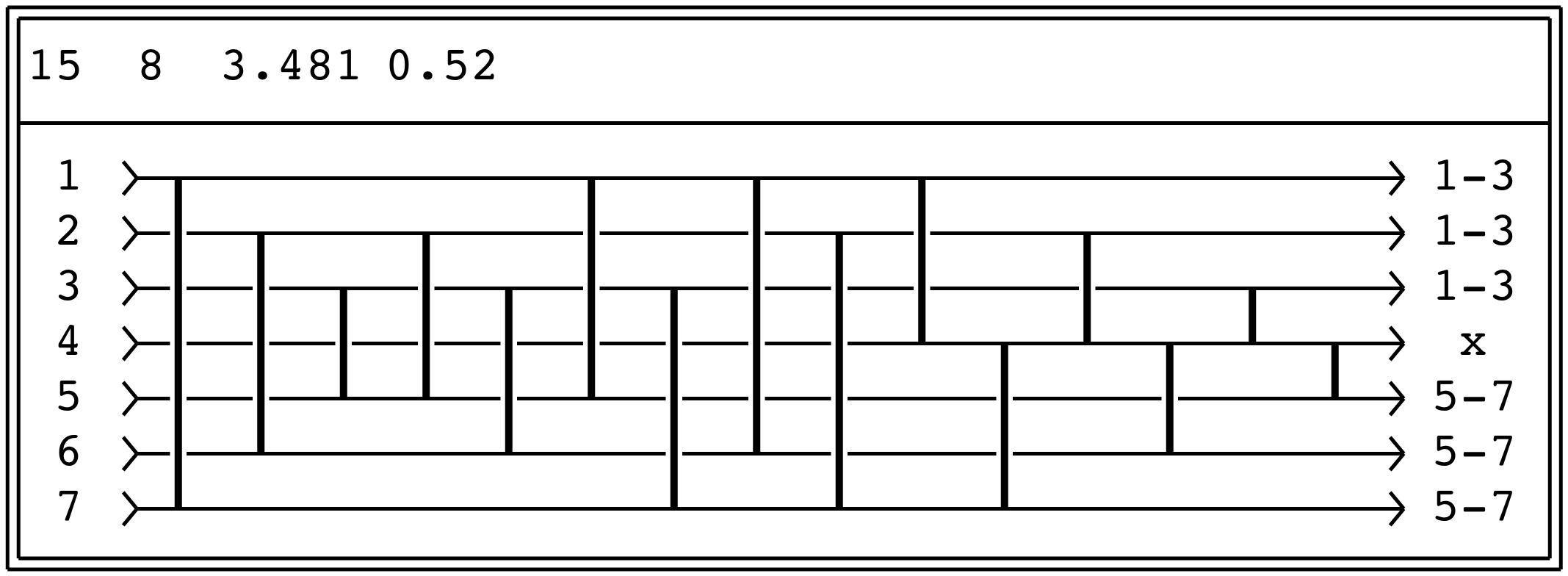}
\end{center}
 \caption{\small
 A network to determine the median of $7$ elements, with a low average of
 $3.481$ exchanges.
 }
\label{fig:21}
\end{figure}

In this section, we describe three networks for $N=7$.
We start with two networks for determining the \emph{median} of $7$ elements.
Figure~\ref{fig:21} is a network for finding the median
of $7$ numbers, requiring a low average of only $3.481$ exchanges.

\begin{figure}[H]
\begin{center}
\includegraphics[width=0.6\textwidth]{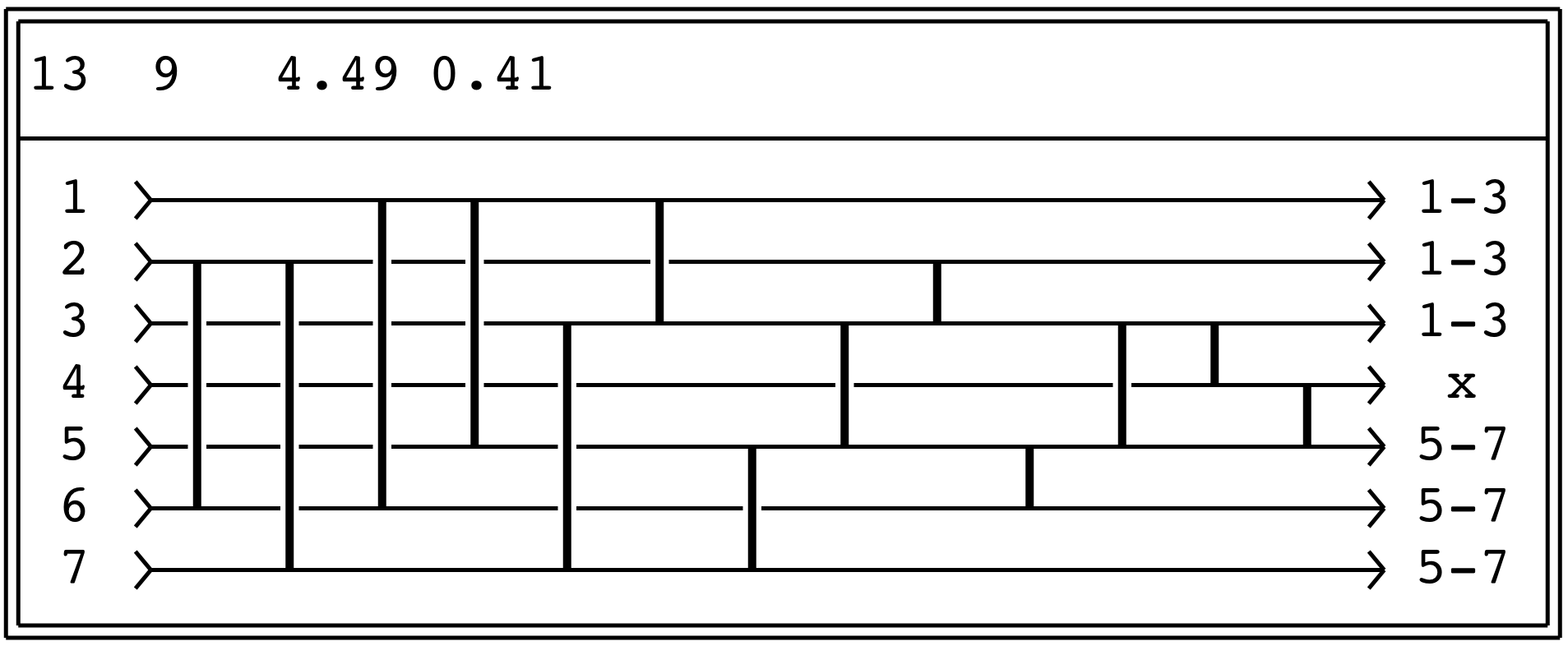}
\end{center}
 \caption{\small
 A network to determine the median of $7$ elements, with a low number of 13
 links.
 }
\label{fig:22}
\end{figure}

The network in Figure~\ref{fig:21} has 15 links.
The network in Figure~\ref{fig:22} finds the median with only 13 links,
albeit increasing the average
of exchanges to $4.49$.

We now turn to \emph{fully sorting} 7 elements.
The network in Figure~\ref{fig:23} has a low average of $5.271$ exchanges:
\begin{figure}[H]
\begin{center}
\includegraphics[width=0.8\textwidth]{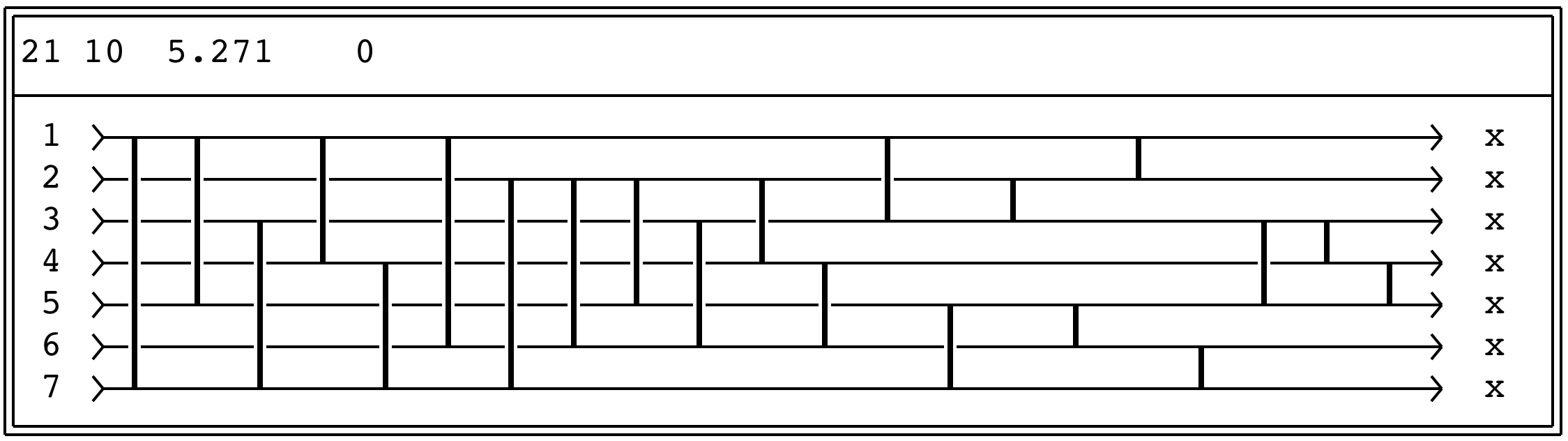}
\end{center}
 \caption{\small
 A network to fully sort $7$ elements, with a low average of $5.271$
 exchanges.
 }
\label{fig:23}
\end{figure}

\subsection{Sorting $N=8$}

We start with the classical \emph{Batcher's odd-even mergesort}
\cite{Batcher}, for sorting 8 elements:
\begin{figure}[H]
\begin{center}
\includegraphics[width=0.8\textwidth]{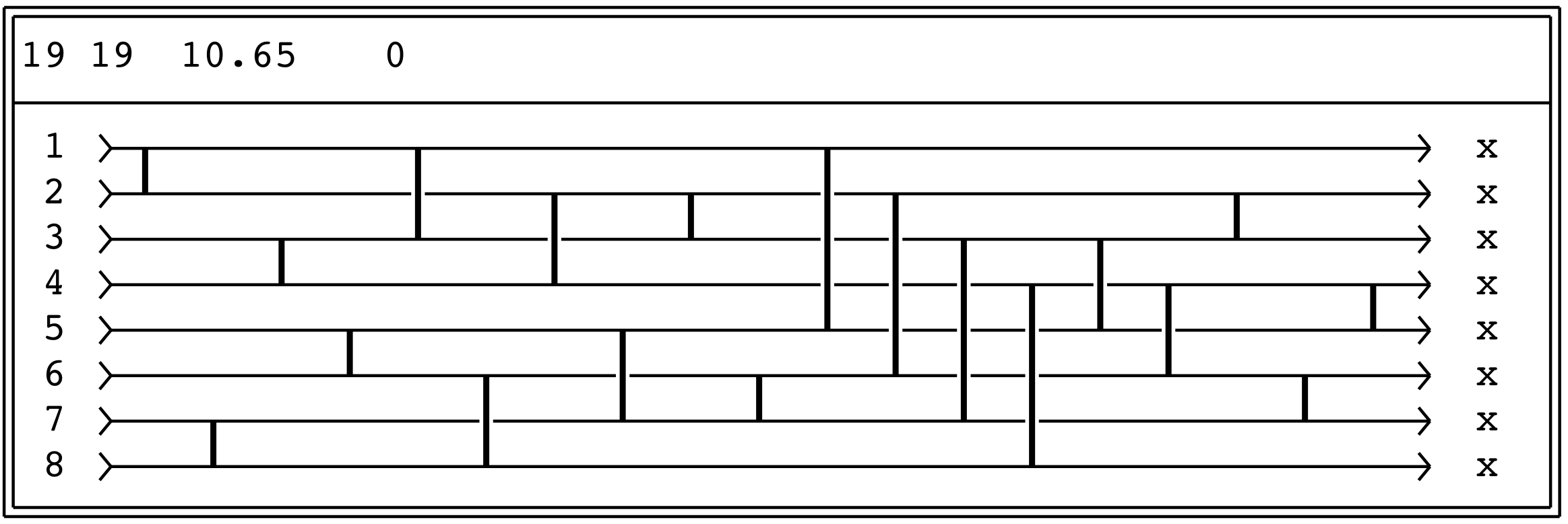}
\end{center}
 \caption{\small
 Batcher's classical odd-even mergesort requires 19 links to sort
 eight elements.
 }
\label{fig:24}
\end{figure}
Batcher's network is known to realize the minimum number of links, 19.
However, we now present two networks with the same number of links, but
with much improved statistics.
First, the network in Figure~\ref{fig:25} has a much improved average number
of exchanges, $7.933$ vs Batcher's $10.65$ and the maximum number
of swaps is only 15 vs Batcher's 19.
\begin{figure}[H]
\begin{center}
\includegraphics[width=0.8\textwidth]{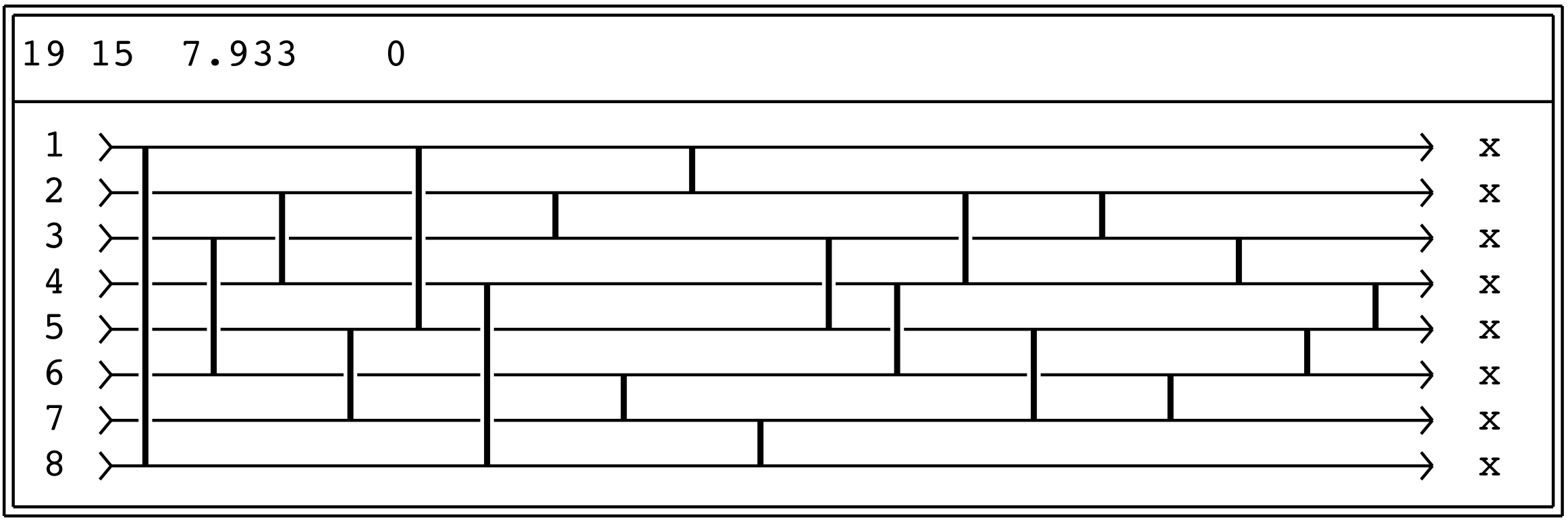}
\end{center}
 \caption{\small
 A network to fully sort $8$ elements, with a low average of $7.933$
 exchanges.
 }
\label{fig:25}
\end{figure}
Second, the network in Figure~\ref{fig:26} has a maximum number of
swaps of only 14 compared to Batcher's 19.
\begin{figure}[H]
\begin{center}
\includegraphics[width=0.8\textwidth]{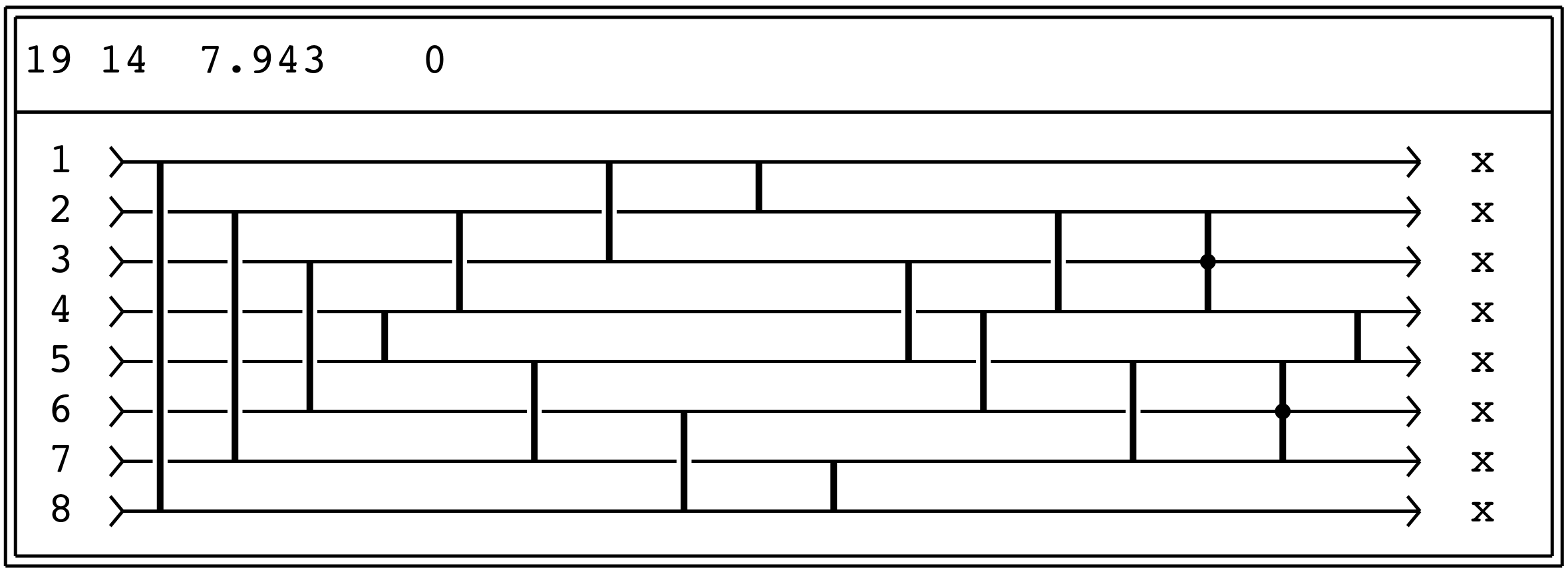}
\end{center}
 \caption{\small
 A network to fully sort $8$ elements, with a low maximum number of swaps
 of $14$.
 }
\label{fig:26}
\end{figure}
We conclude in Figure~\ref{fig:27}
with a network with an even lower maximum number, namely 12,
albeit with 28 links:
\begin{figure}[H]
\begin{center}
\includegraphics[width=0.8\textwidth]{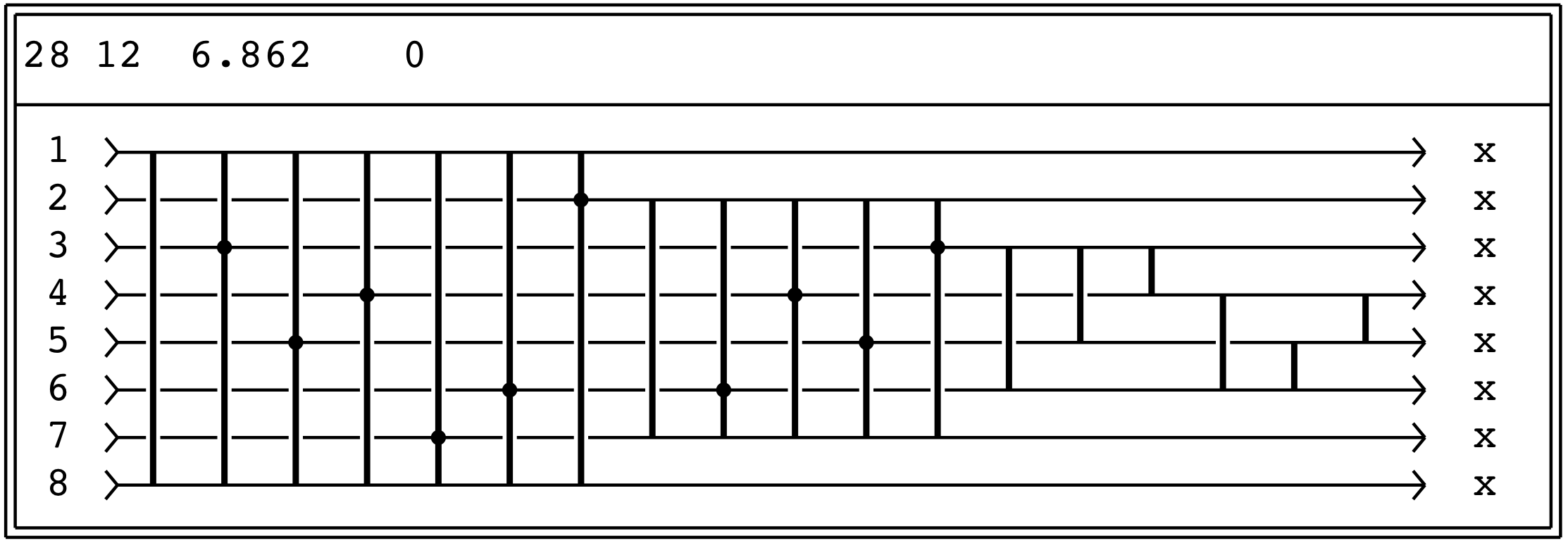}
\end{center}
 \caption{\small
 A network to fully sort $8$ elements, with a low maximum number of swaps
 of $12$.
 }
\label{fig:27}
\end{figure}


\appendixpage

\begin{appendices}

\section{}
\label{app:details}

\begin{center}
\begin{tabular}[t]{@{}lcccc@{}}
\toprule
\multicolumn{4}{c}{Memory contents}\\
\cmidrule{1-4}
Stage 0  &Stage 1 &Stage 2 & Stage 3 & Number of \\
(initial) &(2--3 swap)&(1--2 swap) & (2--3 swap) & swaps\\
\midrule
$(1,2,3)$ &$(1,2,3)$&$(1,2,3)$&$(1,2,3)$ & 0 \\
$(1,3,2)$ &$(1,\mathbf{2},\mathbf{3})$&$(1,2,3)$&$(1,2,3)$ & 1 \\
$(2,1,3)$ &$(2,1,3)$&$(\mathbf{1},\mathbf{2},3)$&$(1,2,3)$ & 1 \\
$(2,3,1)$ &$(2,\mathbf{1},\mathbf{3})$&$(\mathbf{1},\mathbf{2},3)$&$(1,2,3)$ & 2 \\
$(3,1,2)$
&$(3,1,2)$&$(\mathbf{1},\mathbf{3},2)$&$(1,\mathbf{2},\mathbf{3})$ & 2 \\
$(3,2,1)$
&$(3,\mathbf{1},\mathbf{2})$&$(\mathbf{1},\mathbf{3},2)$
&$(1,\mathbf{2},\mathbf{3})$ & 3 \\
\midrule
Number of swaps & 3 & 4 &  2 & 9 \\
\bottomrule
\end{tabular}
\captionof{table}{The behaviour of the network from Figure~\ref{fig:1}(a).}
\label{t:1}
\end{center}%

\begin{center}
\begin{tabular}[t]{@{}lcccc@{}}
\toprule
\multicolumn{4}{c}{Memory contents}\\
\cmidrule{1-4}
Stage 0  &Stage 1 &Stage 2 & Stage 3 & Number of \\
(initial) &(1--3 swap)&(1--2 swap) & (2--3 swap) & swaps\\
\midrule
$(1,2,3)$ &$(1,2,3)         $&$(1,2,3)         $&$(1,2,3)$ & 0 \\
$(1,3,2)$ &$(1,3,2)         $&$(1,3,2)         $&$(1,\mathbf{2},3)$ & 1 \\
$(2,1,3)$ &$(2,1,3)         $&$(\mathbf{1},2,3)$&$(1,2,3)$ & 1 \\
$(2,3,1)$ &$(\mathbf{1},3,2)$&$(1,3,2)         $&$(1,\mathbf{2},3)$ & 2 \\
$(3,1,2)$ &$(\mathbf{2},1,3)$&$(\mathbf{1},2,3)$&$(1,2,3)$ & 2 \\
$(3,2,1)$ &$(\mathbf{1},2,3)$&$(1,2,3)         $&$(1,2,3)$ & 1 \\
\midrule
Number of swaps & 3 & 2 & 2 & 7 \\
\bottomrule
\end{tabular}
\captionof{table}{The behaviour of the network from Figure~\ref{fig:2}(a).}
\label{t:2}
\end{center}%

In Table~\ref{t:1}, we show the $3!=6$ initial memory configurations in
the left column (Stage~0), corresponding to the network in
Figure~\ref{fig:1}(a).
The progress of the sorting network is recorded in each row, with
swaps that occurred highlighted in bold.
This network has 3 stages,
the maximum number of swaps, which occurred for the
initial configuration of $(3,2,1)$ is 3.
The average number of swaps is the total number of swaps ($9$),
divided by the number of configurations ($6$), resulting
in $9/6=1.5$.
Finally, we consider the number of swaps at each stage,
namely 3,4, and 2, divided by the number of configurations,
giving, the ratios $3/6=0.5$, $4/6\approx 0.666667$, and
$2/6\approx 0.333333$, respectively.
This explains all numbers appearing in Figure~\ref{fig:1}(a)
except for the upper right $0$, which represents the quotient
$0/6$, the uncertain positions divided by all positions $3$, thus
representing a measure of unsortedness.

Similar considerations apply to Table~\ref{t:2} which
corresponds to the network in Figure~\ref{fig:2}(a).

\section{}
\label{app:c}

As an example of illustrating the use of the accompanying
code, we used
\begin{verbatim}
./sn -i -o8 18-27-36-45-24=13=12=34=24=234=45- > samplefigure.txt
\end{verbatim}
to generate an ASCII version of Figure~\ref{fig:26}.
The \texttt{=} signifies symmetry, so \verb+24=+ is a shorthand
for \verb+24-57-+. The \texttt{234=} illustrates two (symmetrically placed)
2-ops. The output of this code is a text file that can be post-processed
with code in Appendix~\ref{app:perl} to generate an \texttt{svg} file
which were used to produce the figures in this manuscript.

Below is a listing of the \texttt{C} source code to generate
the networks discussed in this paper.
\lstset{language=C}
\lstset{basicstyle=\small}
\begin{lstlisting}
/////////////////////////////////////////////////////////////////////
// sn.c // factorial style sort-exchange network scanning
//////////////////////////////////////////////////////////////////////

#define VER "2.3"

//	gcc -Ofast sn.c -o sn; mv sn /usr/local/bin
//	gcc -Ofast sn.c -o sn; mv sn /home/awpaeth/bin

#include <stdio.h>
#include <stdlib.h>
#include <string.h>

typedef unsigned long long ULL;

#define RANDX	     16807		// 7 ^ 5

#define BIGPRIME   1000003
#define HLEN	    300007

#define HALFTHRESH	23
#define MAXORD		12
#define MAXL		32		// max number of links

#define Max(a,b) ((a)>(b)?(a):(b))

// if an n-op is otherwise white, make it yellow.
// n-ops are not counting total links properly

// no n-op collision detection in snbin() mode
// snfact() needs (also) to come up with the # of live links.
// broken flag on collision may alwo wish to color links red

// good for 1-32: A-W, use malloc, bit indicies
// more extensive interative mode

// hash array				 HLEN BIGPRIME
// 101,307,1009,3001,10007,30011,100003,300007,1000003,3000017,
// 10000019,30000001,100000007,300000007,1000000007,3000000019

// global structures

ULL hash[HLEN];
int hhit[HLEN];
int hsiz;

// global flags

int xflag = 0;
int wind  = 0;
int rflag = 0;
int ascii = 0;
int bw    = 0;

// global stats (post run)

ULL	gmax;		// max swaps
double	gavg;		// avg swaps
char	gpat[MAXL+1];	// match pattern, e.g. "...X...", ".;,,...", etc.
double  gent[MAXL];	// 0 -> sorted, ~1 -> unsorted
ULL	gtos;		// # of trials conducted (2^n or n! unless monte carlo)
int	glo[MAXL];
int	ghi[MAXL];

// misc globals

int order = 0;

void snfact(), display();
char *getenv();

#define LINKS 600
#define Bt(x) (1ULL<<(x))

ULL bits(ULL x)    { ULL c = 0; while(x) { c++; x &= x-1; } return c; }
int msb(int  x)    { while(bits(x) > 1) x &= x-1; return x; }
int lsb(int  x)    { return (x & (x-1)) ^ x; }
int sortbit(int a) { return ((Bt(bits(a)))-1); }

int bitpos(int x)
    {
    return (
	((x&0xFFFF0000) ? 16:0) +
	((x&0xFF00FF00) ?  8:0) +
	((x&0xF0F0F0F0) ?  4:0) +
	((x&0xCCCCCCCC) ?  2:0) +
	((x&0xAAAAAAAA) ?  1:0) );
    }

void stir(int x, int pbuf[])
    {
    int i, l, r, t;
    for (i = 0; i < order; i++) pbuf[i] = i;
    for (i = order; i > 1; i--)
	{
	r = i - 1; l = x % i; x /= i;
	if (l != r) { t = pbuf[l]; pbuf[l] = pbuf[r]; pbuf[r] = t; }
	}
    }

// hash operations

void hinit()
    {
    int idx;
    hsiz = 0;
    for (idx=0; idx<HLEN; idx++) hhit[idx] = hash[idx] = 0;
    }

int hget(ULL el)
    {
    int rem, idx;
    idx = rem = el % HLEN;
    if (!rem) rem++;
    do  {
	if (!hhit[idx]) return -1;
	if (hash[idx] == el) return idx;
	idx += rem;
	} while (idx != rem);
    return -1;
    }

void hput(ULL el)
    {
    int rem, idx;
    int t = el;
    if (hsiz >= HLEN) printf("hash full!"), exit(0);
    idx = hget(el);
    if (idx >= 0) hhit[idx]++;		// present: tally up
    else
	{				// new: record it
	idx = rem = el % HLEN;
	if (!rem) rem++;
	while (hhit[idx]) idx += rem;	// scan 0x or more for hole
	hash[idx] = el;
	hhit[idx] = 1;
	hsiz++;
	}
    }

void main(int argc, char *argv[])
    {
    int agc = 0; char *agv[50];
    int i, *p, pat[LINKS], link[LINKS], alive, dead;
    char l, *pstr, *pend, *psave;
    int mask, omask;
    int usage, inter, sholnk, echo;
    int canon, noncanon, silent, dflag;
    char *snenv, title[80];
    usage = inter = sholnk = echo = silent = dflag = 0;
    canon = noncanon = 0;
// set up shadow argc/argv (agc/agv)
    agv[agc++] = strdup(argv[0]);
// check for extra args in SN
    if ((snenv = getenv("SN")) && snenv[0])
	{
	char s[2000];
	sprintf(s, "%s", snenv);
	for (i = 0; i<strlen(s); i++)   if (s[i]=='\t') s[i] = ' ';
	for (i = strlen(s)-1; i>0; i--) if (s[i]==' ' ) s[i]='\0'; else break;
	pstr = s;
	while (*pstr)
	    {
	    while (*pstr == ' ') pstr++;
	    pend = pstr;
	    while (*pend && (*pend != ' ')) pend++; *pend = '\0';
	    agv[agc++] = strdup(pstr);
	    pstr = pend+1;
	    }
	}
    for (i = 1; i < argc; i++) agv[agc++] = strdup(argv[i]);
    pstr = 0;

// check args
    for (i = 1; i < agc; i++)
	{
	if (agv[i][0] == '-') switch(agv[i][1])
	    {
    case 'h': dohelp(agv[0]); break;	// HELP text
    case 'a': ascii  = 1; break;	// strictly ascii (vs UNICODE) glyphs
    case 'b': bw     = 1; break;	// b/w graphics, no color
    case 'c': canon  = 1; break;	// canonical patterns, else exit
    case 'd': dflag  = 1; break;	// dead flag: exit on red links
    case 'e': echo   = 1; break;	// echo pattern string
    case 'i': inter  = 1; break;	// interactive mode
    case 'l': sholnk = 1; break;	// link stats summary
    case 'o': order  = atoi(&agv[i][2]); break;	// force order eg 'sn -o8 17='
    case 'r': rflag  = 1; break;	// randomize: partial search space
    case 's': silent = 1; break;	// silent: no output, just return codes
    case 'w': wind   = 1; break;	// window to scan n-ops
    case 'x': xflag  = 1; break;	// xtra xperimental flag
    default:  fprintf(stderr, "unknown flag '%c'\n", agv[i][1]), exit(1);
	    }
	else { if (pstr) usage = 1; pstr = &agv[i][0]; }
	}
    if ((agc == 1) || usage || !pstr)
	fprintf(stderr, "USAGE %s: {-i -b -r -o} 17-26=\n", agv[0]), exit(-1);

// switch overrides
    if (ascii || bw) inter = 1;

// we are live
    for (i = 0; i < MAXL+1; i++) gpat[i] = 0;
    for (i = 0; i < MAXL;   i++) gent[i] = 0;
    psave = pstr;
    p = &pat[0];

// find order
    if (!order) while (l =*pstr++)
	{
	int n = 0;
	if (l >= 'a') l += 'A' - 'a';
	if (('1' <= l) && (l <= '9')) n = l - '0';
	if (('A' <= l) && (l <= 'V')) n = l - 'A' + 10;
	if (n > order) order = n;
	}
    pstr = psave;

// possible warnig
    if ((order<=9) && rflag) fprintf(stderr, "warning: -r increased cost\n");
    if ((order>13) &&!rflag) fprintf(stderr,"13!>2^32: -r essential\n"),exit(1);

// get all links
    mask = 0;
    while (l = *pstr++)
	{
	if (l >= 'a') l += 'A' - 'a';
	if ((l == '-') || (l == '='))
	    {
	    if (mask==0) fprintf(stderr,"bad leading char '%c'\n", l),exit(-2);
	    if (bits(mask)==1) fprintf(stderr,"only one char\n"   ),  exit(-2);

// canonicity check
	    if ( canon &
		((mask & omask) == 0) && (lsb(omask) > lsb(mask)) ) exit(1);
	    omask = mask;

	    if (l == '-') { *p++ = mask; mask = 0; }
	    else
		{
		int mask2 = 0;
		for (i=0; i<order; i++) if (mask&Bt(i)) mask2 |= Bt(order-i-1);
		if ((mask & mask2) && (mask == mask2)) *p++ = mask | mask2;
		else { *p++ = mask; *p++ = mask2; }
		}
	    mask = 0;
	    }
	else if (('1' <= l) && (l <= '9')) mask |= Bt(l - '1');
	else if (('A' <= l) && (l <= 'V')) mask |= Bt(l + 9 - 'A');
	else fprintf(stderr,"bad character '%c'\n", l), exit(-2);
	}
    *p = 0;
    if (mask) fprintf(stderr,"no trailing '-' or '='\n"), exit(-2);
    if (noncanon) exit(1);	// non-canonical if links out of order

//////////////////////////////////
// GENTLEMEN, START YOUR ENGINES
   snfact(pat, link);
//////////////////////////////////

// check for dead links
    alive = dead = 0;
    p = &pat[0]; i = 0;
    while (mask = *p++) if (link[i++]) alive += bits(mask)-1; else dead++;
    if (dflag && dead) exit(2);	// non-canonical if dead links appear

// display results
    if (!silent)
	{
	title[79] = '\0';
	double avent=0; for (i=0; i<order; i++) avent += gent[i]; avent/=order;
	sprintf(title, "%d %2Ld %6.4g %4.2g  %s  %s",
	    alive, gmax, gavg, avent, (inter?"":gpat), (echo?psave:""));
	}
    if (inter) display(pat, link, title);
    else if (!silent) printf("%s\n", title);
// hack in link stats
    i = 0;
    p = &pat[0];
    if (!silent && sholnk)
	while (*p++) printf("%d %g\n", i+1, (double)(link[i])/gtos), i++;
// quit
    exit(0);
    }

void snfact(int pin[], int plink[])
    {
    int gfit[64];		// +1 when rank and sorted element match
    int *p;
    int perm[MAXL];
    int i, lx, k, t, l, r, q;
    ULL ifact, swaps, maxswaps, avgswaps;
    int tmp, swapped;
    double entro = 0.0;

// clear link stats
    i = 0; p = &pin[0]; while (*p++) plink[i++] = 0;

// run all patterns
    ifact = 1; for (i=1; i<=order; i++) ifact *= i;
    if (rflag) ifact = BIGPRIME;
    for (i=0; i<order; i++) gfit[i] =   0;
    for (i=0; i< MAXL; i++) glo[i]  = 100;
    for (i=0; i< MAXL; i++) ghi[i]  =  -1;
    maxswaps = avgswaps = 0;
    for (i=0; i<ifact; i++)
	{
	int match = 0;
	swaps = 0;
	p = &pin[0];
	lx = 0;
	if (rflag) stir((int)((((long long)i)*RANDX)%BIGPRIME), perm);
	    else   stir(i, perm);
	while (t = *p++)
	    {
	    swapped = 0;
	    l = bitpos(lsb(t));
	    while (t &= t-1)
		{
		r = l;
		l = bitpos(lsb(t));
		if (perm[l] < perm[r])
		    {
		    tmp = perm[l]; perm[l] = perm[r]; perm[r] = tmp;
		    if (swapped)
			{
			fprintf(stderr,"link #%d (%d,%d) collision\npattern: ",
			    lx, r+1, l+1);
			stir(i, perm);		// reconstitute orig. pattern
			for (i=0; i<order; i++)
			    fprintf(stderr,"%d ", perm[i]+1);
			fprintf(stderr,"\n");
			exit(2);
			}
		    plink[lx]++;
		    swaps++;
		    swapped = 1;
		    }
		else swapped = 0;
		lx++;
		}
	    }
// record proper positions
	for (k=0; k<order; k++)
	    {
	    q = perm[k];
	    if (q == k) gfit[k]++;
	    if (q < glo[k]) glo[k] = q;
	    if (q > ghi[k]) ghi[k] = q;
	    }

// and stats
	if (swaps > maxswaps) maxswaps = swaps;
	avgswaps += swaps;
	}
// and marker string
    for (i=0; i<order; i++)
	{
	gpat[i] = (gfit[i]==ifact)?(((k*2+1) == order)?'X':'x') : '.';
	gent[i] = (double) (ifact-gfit[i]) / ifact;
	}

// record results
    gmax = maxswaps;
    gavg = (double) avgswaps / ifact;
    gtos = ifact;
    }


///////////////////////////////////////////////////////////////
// display globals
///////////////////////////////////////////////////////////////

char *cred = "\033[30;31m";
char *cgrn = "\033[30;32m";
char *cyel = "\033[30;33m";
char *cblu = "\033[30;34m";
char *cpur = "\033[30;35m";
char *ccyn = "\033[30;36m";
char *cltw = "\033[30;37m";	// normal white
char *cwhi = "\033[30;39m";	// bright white
char *cbld = "\033[1m";
char *cnrm = "\033[m";

#define chorz	cltw		// horizontal rank lines
#define cdead	cred		// dead links
#define cbord	ccyn		// text annotations

void display(int pt[], int lk[], char *anno)
    {
    int i, j, *p, lnum, lmin, lpen;
    char *boxh, *boxv, *boxHu, *boxHd, *boxHv;
    char *arrR, *boxCc;
    char *boxH, *boxV, *boxDR, *boxDL, *boxUL, *boxUR, *boxVr, *boxVl;
    int dsplen, tmargin, bmargin, bwidth;

// double lines, corners
    arrR   = "\342\206\222";				//	->  (arrow rt.)
// double lines, corners
    boxH  = "\342\225\220"; boxV  = "\342\225\221";	//	=	 ||
    boxUL = "\342\225\232"; boxUR = "\342\225\235";	//	||_	_||
    boxDR = "\342\225\224"; boxDL = "\342\225\227";	//	||-	-||

// single / single, single+double lines
    boxCc = "\342\227\213"; // <or> \342\227\257	//	o (big)
    boxCc = "\342\217\200";				//	o (line thru)
//  boxCc = "\342\247\263";				//	o (Alternate)
    boxCc = "\342\215\277";				//	* (barred)

    boxh  = "\342\224\200"; boxv  = "\342\224\202";	//	-	 |
    boxHu = "\342\224\254"; boxHd = "\342\224\264";	//	perp    tee
    boxVr = "\342\225\237"; boxVl = "\342\225\242";	//	|-	-|
    boxHv = "\342\224\277"; // \274 aka boxvh stock +	//	+ (w/ thick)


    boxHv = boxCc;					// center '+' now a 'o'
    boxv  = "\342\224\203";
    boxHu = "\342\224\260";
    boxHd = "\342\224\270";

    if (ascii)
	{
	arrR  = ">";
	boxHv = "o";					// center '+' now a 'o'
	boxH  = "="; boxV  = "|";
	boxh  = "-"; boxv  = "|";
	boxUL = boxUR = boxDR = boxDL = "+";
	boxHu = boxHd = boxVr = boxVl = "+";
	}

    if (bw)
	{
	cred = cgrn = cyel = cblu = cpur = ccyn = cltw = cwhi = cbld = cnrm;
	}

    lnum = 0; p = &pt[0]; while(*p++) lnum++;
    lpen = lmin = gtos;
    for (j = 0; j < lnum; j++) if (lk[j] && (lk[j] < lmin)) lmin = lk[j];
    for (j = 0; j < lnum; j++) if (lk[j] && (lk[j] > lmin) && (lk[j] < lpen))
								lpen = lk[j];
    if ((lpen*2) >= gtos) { lpen = lmin; lmin = 0; }

// top line
    dsplen = ((lnum<HALFTHRESH) ? 3:1)*lnum+10;	// 10 for body line margin junk
    tmargin = Max(dsplen, strlen(anno)) - strlen(anno);
    bmargin = Max(dsplen, strlen(anno)) - dsplen;
    bwidth = bmargin + dsplen;

    printf("%s", cbord);
    printf("%s",boxDR);
	 for (j=0; j<bwidth; j++) printf("%s",boxH); printf("%s\n",boxDL);
    printf("%s",boxV); printf("%s", anno);
	 for (j=0; j < tmargin; j++) printf(" ");
	 printf("%s\n", boxV);
    printf("%s",boxVr);
	 for (j=0; j<bwidth; j++) printf("%s",boxh); printf("%s\n",boxVl);
    printf("%s", cnrm);

    for (i=0; i<order; i++)
	{
	printf("%s%s%s %c >", cbord,boxV,cnrm, (i>8)?i-9+'A':i+'1');
	p = &pt[0];
	for (j=0; j < lnum; j++)
	    {
	    char *s, t[20];
	    int l, r;
	    int m = *p++;
	    r = bitpos(lsb(m));
	    l = bitpos(msb(m));
	    s = boxh;
	    sprintf(t,"%s%s%s", chorz, s, chorz);
// feature control
	    if ((r <= i) && (i <= l))
		{
		s = boxv;
		if (l == i) s = boxHd;
		if (r == i) s = boxHu;
		if ((i != l) && (i != r) && (m & Bt(i))) s = boxHv;
		if ((lk[j]*2) > gtos)	sprintf(t,"%s%s%s", cyel, s, cnrm);
		else if (lk[j] == 0)	sprintf(t,"%s%s%s", cdead,s, cnrm);
		else if (lk[j] == lmin)	sprintf(t,"%s%s%s", cpur, s, cnrm);
		else if (lk[j] == lpen)	sprintf(t,"%s%s%s", cblu, s, cnrm);
		else			sprintf(t,"%s%s%s", cwhi, s, cnrm);
		}
	    if (lnum < HALFTHRESH) printf("%s%s%s", chorz,boxh,chorz);
	    printf("%s", t);
	    if (lnum < HALFTHRESH) printf("%s%s%s", chorz,boxh,chorz);
	    }
	if (glo[i] != ghi[i])
	    printf("> %c-%c ",
		 ((glo[i]>=9) ? glo[i]-9+'A' : glo[i]+'1'),
		 ((ghi[i]>=9) ? ghi[i]-9+'A' : ghi[i]+'1') );
	else printf("%s  %c  ", arrR, gpat[i]);
	for (j = 0; j < bmargin; j++) printf(" ");
	printf("%s%s%s\n", cbord,boxV,cnrm);
	}

// bottom line
    printf("%s", cbord);
    printf("%s",boxUL);
	for (j=0; j<bwidth; j++) printf("%s",boxH); printf("%s\n",boxUR);
    printf("%s", cnrm);
    }

dohelp(char *prog)
    {
    printf("%s ver %s\n%s\n",prog, VER,"\
    'h': this text\n\
    'a': strictly ascii (vs UNICODE) glyphs\n\
    'b': graphics in b/w, no color\n\
    'c': canonical patterns, else exit\n\
    'd': dead flag: exit on red links\n\
    'e': echo pattern string\n\
    'i': interactive mode\n\
    'l': link stats summary\n\
    'o': force order eg 'sn -o8 17='\n\
    'r': randomize: partial search space\n\
    's': silent: no output, just return codes\n\
    'w': (future) window to scan n-ops\n\
    'x': (future) xtra xperimental flag\n\
frequently used flags may be placed in env. var SN");

    exit(1);
    }
\end{lstlisting}

\section{}
\label{app:perl}

The output file \texttt{samplefigure.txt} described in Appendix~\ref{app:c}
can be postprocessed with
\begin{verbatim}
./fix samplefigure.txt > samplefigure.svg
\end{verbatim}
to generate Figure~\ref{fig:26}.

Below is a listing of the \texttt{perl} code to generate
the networks discussed in this paper.
\lstset{language=Perl}
\lstset{basicstyle=\small}
\begin{lstlisting}
#!/usr/bin/perl

# fix.pl - convert UTF8 graphics (box characters) into SVG/HTML

$xscale =  7.5;
$yscale = 15.0;

$mscale = $xscale; if ($yscale < $mscale) { $mscale = $yscale; }

#	( )
#	@ $

## unused:	<  %  ^  &

foreach (<>)
    {
    chomp;

# protect hyphens in intervals (convert to underscores)
    s/([0-9])-([0-9])/$1_$2/g;

# map UTC8 box formers
    s/\e\[30;3.m//g;
    s/\e\[m//g;
    s/\342\206\222/>/g;		# arrow right
    s/\342\215\277/*/g;		# vert dot 2-op
    s/\342\225\221/#/g;		# double col
    s/\342\225\242/{/g;		# double col LHS Tee
    s/\342\225\237/}/g;		# double col RHS Tee
    s/\342\225\220/=/g;		# double row
    s/\342\225\224/;/g;		# NW double corner
    s/\342\225\227/:/g;		# NE double corner
    s/\342\225\232/`/g;		# SW double corner
    s/\342\225\235/'/g;		# SE double corner
    s/\342\224\260/~/g;		# top Tee
    s/\342\224\270/!/g;		# bot Tee (perp)
    s/\342\224\200/-/g;		# single row
    s/\342\224\203/|/g;		# single col

    s/>-/?-/g;			# arrow (start)
# store new row
    $rbuf[$rlim++] = $_;
    }

# convert chars into stub lines
    gron();
# GRAPHICS conversion
    for ($row = 0; $row < $rlim; $row++)	# GRAPHICS loop
	{
	$col = 0;
	for $c (split '', $rbuf[$row])
	    {
#	. * . * .	. * . * .	. * . * .	. . . . .
#	. * . * .	. * . * .	. * . * .	* * * * *
#    #	. * . * .    {	* * . * .    }	. * . * *    =	. . . . .
#	. * . * .	. * . * .	. * . * .	* * * * *
#	. * . * .	. * . * .	. * . * .	. . . . .

	    if ($c eq '#') { v(.3,0,1); v(.7,0,1); }
	    if ($c eq '{') { v(.3,0,1); v(.7,0,1); h(.0, .5,.3); }
	    if ($c eq '}') { v(.3,0,1); v(.7,0,1); h(.7, .5,.3); }
	    if ($c eq '=') { h(0,.4,1); h(0,.6,1); }

#	. . . . .	. . . . .	. * . * .	. * . * .
#	. * * * *	* * * * .	. * . * *	* * . * .
#    ;	. * . . .    :	. . . * .    `	. * . . .    '	. . . * .
#	. * . * *	* * . * .	. * * * *	* * * * .
#	. * . * .	. * . * .	. . . . .	. . . . .

	    if ($c eq ';') { v(.3,.4,.6);v(.7,.6,.4);h(.3,.4,.7); h(.7,.6,.3);}
	    if ($c eq ':') { v(.7,.4,.6);v(.3,.6,.4);h(0.,.4,.7); h(0.,.6,.3);}
	    if ($c eq '`') { v(.3,0.,.6);v(.7,0.,.4);h(.3,.6,.7); h(.7,.4,.3);}
	    if ($c eq '\''){ v(.3,0.,.4);v(.7,0.,.6);h(0.,.4,.3); h(0.,.6,.7);}

#	. . * . .	. . . . .	. . . . .	. . * . .
#	. . * . .	. . . . .	. . . . .	. . * . .
#    !	* * * * *    ~	* * * * *    -	* * * * *    |	* . * . *
#	. . . . .	. . * . .	. . . . .	. . * . .
#	. . . . .	. . * . .	. . . . .	. . * . .

	    if ($c eq '!') { h(0.,.5,1); vv(.5,0.,.5); }
	    if ($c eq '~') { h(0.,.5,1); vv(.5,.5,.5); }
	    if ($c eq '-') { h(0.,.5,1); }
	    if ($c eq '|') { vv(.5,0.,1); h(0.,.5,.2); h(.8,.5,.2); }

#	. . . . .	. . * . .	. . * . .	. . . . .
#	. . * . .	. . . * .	. . . * .	. . . . .
#    *	. * * * .    >	* * * * *    ?	. . . . *     	. . . . .
#	. . * . .	. . . * .	. . . * .	. . . . .
#	. . . . .	. . * . .	. . * . .	. . . . .


	    if ($c eq '*') { v(.5,0.,1); h(0.,.5,.38); h(.62,.5,.38);
								c(.5,.5,.25); }
	    if ($c eq '>') { h(0,.5,1); d(1,.5,.5,.2,1); d(1,.5,.5,.8,1); }
	    if ($c eq '?') {            d(1,.5,.5,.2,1); d(1,.5,.5,.8,1); }

#	. . . . .	. . . . *	. . . . .	* . . . .
#	. . . . .	. . . * .	. . . . .	. * . . .
#    (	* * * . .    )	* * * . .    @	. . * * *    $	. . * * *
#	. . . * .	. . . . .	. * . . .	. . . . .
#	. . . . *	. . . . .	* . . . .	. . . . .

	    if ($c eq '(') {
printf "<path d=\"m %d %d\n", $xscale*$col, $yscale*$row;
print  "m 0 7.5 l 6.5 6.5\" stroke=\"black\" fill=\"transparent\"\/>\n"; }
	    if ($c eq ')') {
printf "<path d=\"m %d %d\n", $xscale*$col, $yscale*$row;
print  "m 2 16 l 6.5 6.5\" stroke=\"black\" fill=\"transparent\"\/>\n";}
	    if ($c eq '@') {
printf "<path d=\"m %d %d\n", $xscale*$col, $yscale*$row;
print  "m 0 7.5 l 7.5 -7.5\" stroke=\"black\" fill=\"transparent\"\/>\n"; }
	    if ($c eq '$') {
printf "<path d=\"m %d %d\n", $xscale*$col, $yscale*$row;
print  "m 0 0.5 l 8 -8\" stroke=\"black\" fill=\"transparent\"\/>\n";}
	    $col++;
	    }
	}
# CHAR conversion
    for ($row = 0; $row < $rlim; $row++)	# CHAR loop
	{
	@cbuf = split //, $rbuf[$row];
	$on = 0;
	for ($col = 0; $col <= $#cbuf; $col++)
	    {
	    $c = $cbuf[$col];
	    $f = (('a' le lc($c)) && (lc($c) le 'z')) ||
		 (('0' le lc($c)) && (lc($c) le '9')) ||
		 ($c eq '_') ||
		 ($c eq '.');
	    if ($f)
		{
		if ($c eq '_') { $c = '-'; }
		if (!$on) { $on = 1; $buf = ""; $crow = $row; $ccol = $col; }
		if ( $on) { $buf .= $c; }
		}
	    if (!$f || ($col == $#cbuf))
		{
		if ( $on) { $on = 0; t($buf, $ccol, $crow); }
		if (!$on) {  }
		}
	    }
	}
# finish
    groff();
    exit 0;

sub h
    {
    ($x0, $y0, $w) = @_;
    d($x0, $y0, $x0 + $w, $y0, 1);
    }

sub v
    {
    ($x0, $y0, $h) = @_;
    d($x0, $y0, $x0, $y0 + $h, 1);
    }

sub vv
    {
    ($x0, $y0, $h) = @_;
    d($x0, $y0, $x0, $y0 + $h, 2);	# increased thickness
    }

sub c
    {
    ($x0, $y0, $rad) = @_;
    $x0 += $col; $y0 += $row;
    $x0 *= $xscale; $y0 *= $yscale; $rad *= $mscale;
    printf "<circle cx=%g cy=%g r=%g stroke=\"black\"/>\n",
		$x0, $y0, $rad;
    }

sub t
    {
    $n = 7;
    @fixed = (	"Fira Mono", "DejaVu Sans Mono", "Menlo", "Consolas",
		"Liberation Mono", "Monaco", "Lucida Console", "monospace",
 		"Nimbus Mono", "georgia" );
    ($txt, $x0, $y0) = @_;
    $y0 += 0.7; $x0 -= 0.0;	# font mbb offsets
    $x0 *= $xscale; $y0 *= $yscale; $rad *= $mscale;
    $font = $fixed[$n];
    printf
    "<text x=\"%g\" y=\"%g\" fill=\"black\" font-family=\"$font\">%s</text>\n",
		$x0, $y0, $txt;
    }

sub d
    {
    ($xs, $ys, $xe, $ye, $th) = @_;
    $xs += $col; $ys += $row;
    $xe += $col; $ye += $row;
    $xs *= $xscale; $ys *= $yscale;
    $xe *= $xscale; $ye *= $yscale;
    push @xs, $xs; push @ys, $ys;
    push @xe, $xe; push @ye, $ye;
    push @th, $th;
    }

sub gron
    {
    print "<svg width=\"700\" height=\"900\" viewPort=\"0 0 700 900\"\n";
    print " version=\"1.1\" xmlns=\"http://www.w3.org/2000/svg\">\n";
#	<rect x=0 y=0 width=200 height=320 fill="linen"/>
    }

sub merge
    {
    for ($i=0; $i<=$#xs; $i++)
	{
        for ($j=$0; $j<=$#xs; $j++)
	    {
	    if ( ($i != $j) && ($xe[$i] == $xs[$j]) && ($ye[$i] == $ys[$j]) &&
( (($xs[$i] == $xe[$i]) && ($xe[$i] == $xs[$j]) && ($xs[$j] == $xe[$j])) ||
  (($ys[$i] == $ye[$i]) && ($ye[$i] == $ys[$j]) && ($ys[$j] == $ye[$j])) ) )
		{
# update ith entry
		$xe[$i] = $xe[$j];
		$ye[$i] = $ye[$j];
# delete jth entry
		$xs[$j] = $xs[$#xs];
		$ys[$j] = $ys[$#ys];
		$xe[$j] = $xe[$#xe];
		$ye[$j] = $ye[$#ye];
		$th[$j] = $th[$#th];
		pop @xs; pop @ys; pop @xe; pop @ye, pop @th;
		return $#xs+1;
		}
	    }
	}
    return 0;
    }

sub groff
    {
# compact the vector list
    while (merge()) {  }
# output  the vector list
    $i = 0;
    for $xs (@xs)
	{
	printf "<line x1=%g y1=%g x2=%g y2=%g stroke=\"black\"",
	    $xs, $ys[$i], $xe[$i], $ye[$i];
	if ($th[$i] != 1)
	    { printf  " stroke-width=\"%g\"", $th[$i]; }
	print  "/>\n";
	$i++;
	}
    print "</svg>\n";
    }
\end{lstlisting}

\section{}
\label{app:AlanEmail}
Finally, we list list the email with Alan Paeth's original draft of
this manuscript.

\fvset{fontsize=\small}
\begin{Verbatim}
From: Alan Paeth <awpaeth@gmail.com>
Date: Wed, 28 Mar 2018 13:48:34 -0700
Subject: conex paper
To: Heinz Bauschke <heinz.bauschke@ubc.ca>
[other recipients suppressed, obvious spelling mistakes corrected,
 formatting of text adjusted for easier reading]

Non-Interfering Concurrent Exchange (NICE) Networks


Synopsis

In studying the statistical frequency of exchange in comparison-exchange (CE)
networks we discover a new elementary form of comparison-exchange which we name
the "2-op". The operation supports concurrent and non-interfering operations of
two traditional CEs upon one shared element. More than merely improving overall
statistical performance, the introduction of NICE (non-interfering CE) networks
lowers long-held bounds in the number of stages required for sorting tasks.
Code-based CEs also benefit from improved average/worst case run time costs.


Motivation

Comparison-based sorts dominate much of modern sorting; in-place methods such as
quicksort are widely employed and well-studied. All seek to minimize the number
of comparisons required. For small numbers of input (N), a sequence of
predetermined comparisons form a decision tree of N! leaves; if minimal, its
height is then ceiling(N log2 N), e.g. for N=5, 7 comparisons suffice to fully
determine the input permutation. To complete the sort a sequence of cyclic
exchanges then reorder the data. These are often simplified into a sequence of
two-element swaps. At N>5 the decision tree's size makes it impractical in
production settings. At N=3 all solutions require a tree having height=3,
leaves=6. The tree show below is optimal in that the two comparison descents
occur with with M[0]<=M[1]<=M[2] or M[0]>=M[1]>=M[2] i.e., termination occurs
early with presorted (ascending or descending) elements:

        if mem[0] <= mem[1] then
             if mem[1] <= mem[2]      then      return // 0 1 2
             else if mem[0] <= mem[2] then      return // 0 2 1
                  else                          return // 2 0 1
        else if mem[1] >= mem[2]      then      return // 2 1 0
             else if mem[0] >= mem[2] then      return // 1 2 0
                  else                          return // 1 0 2

This network is widely quoted [ref]. In the first two steps we either establish
mem[0]<=mem[1] and mem[1]<=mem[2] and gain mem[0]<mem[2] by transitivity (is
sorted). Otherwise, at the first ELSE mem[1] has largest rank; we need merely
disambiguate the ranks of nonadjacent mem[0] and mem[2]. The second half follows
by symmetry.

For larger N we can establish that mem[i]<mem[i+1] in N-1 steps but this set of
comparisons does not lead to an optimal (balanced) decision tree. In the code
seen above exchanges complete the sort. For the respective six leaves, these are
the exchanges { nil;  (1,2);  (1,2),(0,1);  (0,2);  (0,1),(1,2); (0,1); }.

By contrast, compare-exchange sorting networks order an array by performing a
fused compare and conditional-exchange operation. They are ideally suited to
sorting arrays of integers of fixed small size (N = 1..25 typical) where the
decision making overhead of more general methods (tree traversal using
comparison and bifurcation) will diminish or even negate any reduction in total
machine comparisons. An integer compare is typically a single machine
instruction. While compact and efficient, a general methodology for the creation
of optimal fixed compare-exchange networks remains elusive.

Worst, lack of suitable metrics may lead to sub-optimal networks. Below are two
simple networks for N=3. At (a) we recode based on the above algorithm. At (b)
we apply Batcher's even-odd construction for an odd number of elements, applying
central symmetry (of inversion) to the right and left halves.


        >---------+--------->            >---------+----+---->
                  |.67                             |    |.67
        >----+----+----+---->            >----+----|----+---->
             |.5       |.33                   |.5  |.33
        >.---+---------+---->            >----+----+--------->

               3/1.5                            3/1.5

                (a)                              (b)

Exchanges are costly, often at a ratio of 2:1 or 3:1 to a simple comparison. We
have affixed the likelihood of an exchange to each link. Summary statistics give
the maximum and average number of exchanges for the entire network.'

The networks presented so far for N=3 are distinct and require in both cases
three stages, comparisons and exchanges. Both also demonstrate that a full sort
network will include all possible length=1 links as these alone can serve to
reorder permutations of sorted data when merely one adjacent transposition
exists. The symmetry of (b) is compelling: it allows for fully bidirectional
sorting where input and output sides reverse. Unfortunately, it (like (a)) also
exhibits a link in which swapping occurs more often then not.

When a link exchange occurs more than half the time, we may reverse the swap vs
non-swap .67/.33 to .33/.67 by preexchanging its elements. We remove the direct
cost of the exchange by reversing the order of input lines to the left of the
offending link. But all inputs are unsorted so any non-conditional exchange has
no sorting efficacy and may be removed. Applied to links .67 in (a), (b) give

        >----+----+--------->            >----+---------+---->
             |    |.33                        |         |.33
        >----|----+----+---->            >----|----+----+---->
             |.5       |.33                   |.5  |.33
        >----+---------+---->            >----+----+--------->

               2/1.17                           2/1.17

                (c)                              (c')

and the networks, identical under mirror symmetry, substantially improve the
average cost of exchange. More striking is a reduction in maximum exchange,
which was lowered from three to two. This is unexpected and serves as the basis
of non-interfering concurrent exchange. Clearly, the first link may exchange
without restriction. The remaining cost of one exchange must be shared between
the two remaining links: at most one may occur. This in turn implies that the
central element cannot be rewritten by both links, allowing both the upper and
lower link concurrent execution.

We now create a new fused dual-link element, which we name a '2-op' (borrowing
from mathematical nomenclature used for such occurrences), depicted below:

                        >----+----+---->
                             |    |.33
                        >----|----o---->
                             |.50 |.33
                        >----+----+---->

                               (d)

in which the open circle at common link joint indicates non-interference.


Cost Benefits

The 2-op exchange on three elements yields immediate gains in lowering costs of
traditional networks. For example, Paeth [1990] described a median on a 3x3 box,
formed by column, row and (single) diagonal sorting, first conceived as a means
to reuse column sorts when filtering a large raster image.

        +-----+
        |1 2 3| by col          s3(1,4,7) s3(2,5,8) s3(3,6,9)
        |4 5 6| by row          s2(1,2,3) s3(4,5,6) s2(7,8,9)
        |7 8 9| diagonal        s3(3,5,7)
        +-----+

              s3(a,b,c) (original) is s2(a,b) s2(b,c) s2(a,b)
              s3(a,b,c) (reworked) is s2(a,c) s2(a,b) s2(b,c)

Column reuse aside, the method leads to efficient median (19 links) and sorting
(25) on nine element arrays. In 21 links we find minimum, maximum and median,
deletion and inclusion lead both to bare median (19) full sorted (25) forms.


+=========================================================================+
  |21 21  9.771
|

+-------------------------------------------------------------------------+
  | 1 >-+-----+--------------------+-----+---------------------------->  x
|
  | 2 >-|-----|--+-----+-----------+--+--+----------------------------> 2-4
|
  | 3 >-|-----|--|-----|--+-----+-----+-----------------------+-----+-> 2-4
|
  | 4 >-+--+--+--|-----|--|-----|-----------+-----+-----------|-----|-> 2-4
|
  | 5 >----|-----+--+--+--|-----|-----------+--+--+-----------+--+--+->  x
|
  | 6 >----|--------|-----+--+--+--------------+-----------------|----> 5-7
|
  | 7 >----+--------|--------|-----------------------+-----+-----+----> 5-7
|
  | 8 >-------------+--------|-----------------------+--+--+----------> 5-7
|
  | 9 >----------------------+--------------------------+------------->  x
|

+=========================================================================+

        .5 .7 .3 .5 .7 .3 .5 .7 .3 .5 .7 .3 .5 .7 .3 .5 .7 .3 .2 .4 .1

   Notes: .5 = 1/2 .7 = 2/3 .3 = 1/3 .2 = 19/70 (.271) .4 = 5/14 .1 = 1/7

We may arrange the links to minimize the total number of stages as well.
This does not change the overall swap statistics:


  +===================================================+
  |21 21  9.771                                       |
  +---------------------------------------------------+
  | 1 >-+-------------+------+-----+------------>  x  |
  | 2 >-|-+-----------|-+----+--+--+------------> 2-4 |
  | 3 >-|-|-+---------|-|-+-----+-----+-----+---> 2-4 |
  | 4 >-+-|-|--+------+-|-|--+-----+--|-----+---> 2-4 |
  | 5 >---+-|--|-+------+-|--+--+--+--+--+--+--->  x  |
  | 6 >-----+--|-|-+------+-----+--------|------> 5-7 |
  | 7 >--------+-|-|---------+-----+-----+------> 5-7 |
  | 8 >----------+-|---------+--+--+------------> 5-7 |
  | 9 >------------+------------+--------------->  x  |
  |     I      II     III   IV V  VI VII VIII IX      |
  +===================================================+


Exchanges on 21 links occur when presented reverse sorted input, or 9.771 times
on average, giving a relative frequency of 47%. Substitution of sort3 gives


  +=====================================+
  |21 14  7.657 0.35                    |
  +-------------------------------------+
  | 1 >-+------+------+--+-------->  x  |
  | 2 >-|-+----|-+----|--o--+--+--> 1-4 |
  | 3 >-|-|-+--|-|-+--+--+--|--|--> 1-4 |
  | 4 >-|-|-|--o-|-|--+--+--|--|--> 1-3 |
  | 5 >-|-|-|--|-o-|--|--o--|--o-->  x  |
  | 6 >-|-|-|--|-|-o--+--+--|--|-->     |
  | 7 >-+-|-|--+-|-|--+--+--|--|--> 1-4 |
  | 8 >---+-|----+-|--|--o--|--|--> 1-7 |
  | 9 >-----+------+--+--+-------->  x  |
  |       I     II   III IV V  VI       |
  +=====================================+

Full sorting in 25 links adds the four CE elements (2,4) (3,4) (6,8) (6,7); Bare
median in 19 omits the final (1,2) and (8,9) from the 2-ops appearing in stage
IV. Note that these omissions cannot be made directly on the original net
without permutations to allow for median. The final networks are not drawn; the
benefits to both hardware and software implementations are now summarized:

        Min/Max/Med (in 19)             Full Sort (in 25)
          Swaps    Stages                 Swaps   Stages
        avg   max                       avg   max

  old   9.11 19      9                  11.56 25    11
  new   6.99 13      6                   9.44 18     8
  %ch   77%  68%    67%                 72%   82%   73%


Intuition and Extension

To some the lowered statistics seem surprising. Typical reasoning follows: "to
sort three, order any two, the third draws a bye. Bye enters and plays to
establish an overall winner (WLOG, loser); third step ranks the runner's up."
While accurate when elements are chosen freely, as a memory-based method each
element carries a rank implicit in its index. CE sorting ranks all elements,
chosen pairs are not isomorphic. In the case of N=3 we may disambiguate thus:
"Choosing any two at most omits the median, our set is are guaranteed to hold at
least the min or max. Choosing and sorting the outermost elements therefore
guarantees at least on of max or min landing in its final position. A single
swap will place the remaining extrema; both extreme must be treated (a 2-op).
With both extrema placed, median is properly placed by process of elimination".

Students of computer science will recognize (a) as bubble-sort, whose initial
link leaves the maximum element in central position 2/3 of the time (for the
remaining 1/3 it started in proper position and will not move), thereby
encumbering subsequent links the task of floating it into its outer position. By
comparison, the non-interfering method is seen as a selection or insertion sort
whose improved swapping statistics are well-known. It is ironic such study of CE
networks is rare and that the total number of links (or stages) has remained the
sole performance metric for so long.

Conceptually, the longer prefacing element link in (c) serves as a guard in
assuring that no double-exchange occurs, allowing fusion into one 2-op. The
final links might not be unit length, as seen in the improved sort/median on
nine elements. The guard need only protect (operate on) the outer two unshared
elements. As a second and more interesting means of extension, we envision a
3-op of higher order, formed by allowing two adjacent 2-ops to execute
concurrently. We explore the latter now.


Sorting N=4

A trial network (e) overlaps two optimal 3-sorts, sharing a central element,

                >--+-------+--->         >--+-------+------->
                   |       |                |       |
                >--|---+---o--->         >--|---+---o---+--->
                   |   |   |                |   |   |   |
                >--+---|---o--->         >--+---|---o---+--->
                       |   |                    |   |
                >------+---+--->         >------+---+------->

                       (e)                        (f)

whose sorting statistics and testing reveal a maximum swap of four elements,
occurring only with the central 3-op not active, as desired. Unfortunately, this
network fails to fully sort in five steps: 1/3 of the time the central elements
are transposed, requiring a resort to the six link for seen in (f). This is both
undesirable and surprising: the final link seemingly replicates the work
concluded immediately its left. There are two remedies to this. Both rely on the
fact that the central link in the 3-op might not execute, yet its operation is
essential to attain a full sort.

Our first solution breaks the 3-op into its constituent links. Since they do not
interfere, we are afforded any ordering of its components. We first choose to
place the link to its right, giving rise to redundant right-most central links
(not shown). We retain one only, giving the familiar solution (g), sans N-ops.

The second (h) recognizes that the additional link is necessary in the case the
3-op is fully active (two swaps, outer), its actions then juxtaposing two (as
yet unordered) elements against each other into adjacent central position. As
before we guard against interference by employing a prefacing link whose two
endpoints are the outer endpoints (elements) of the two links it protects:

                >--+-------+------->     >--+---+-------+--->
                   |       |.5              |   |.33    |.25
                >--|---+---+---+--->     >--|---|---+---o--->
                   |.5 |       |.33         |   |   |   |.33
                >--+---|---+---+--->     >--|---+---|---o--->
                       |.5 |.5              |.5     |.33|.25
                >------+---+------->     >--+-------+---+--->

                max=5     avg=2.33       max=4      avg=2.0

                       (g)                        (h)

The second solution is in fact the O(n^2) pair-wise comparison of all elements,
taken in descending order, which necessarily ranks all elements. While not a
viable solution for large N, for N=4 it sits close to the theoretical minimum (5
links) and rewards the implementor with surprisingly good statistics for average
swaps (2) while nonetheless executing in at most three hardware stages.

Large networks whose size N is a power of two, have many prefacing links whose
sorting statistics show a 50/50 probability of exchange. The transposition
operator may be applied to any such link (and to any implicated links lying to
its left) without changes to the average sorting statistics. The topological
change may instead succeed in reducing the maximum number of exchanges.

Applied to the third or fourth link (but not both) seen in stage (g) II yields

                >--+-------+------->
                   |       |.5
                >--|---+---+---+--->
                   |   |.5     |.33
                >--|---+---+---+--->
                   |.5     |.5
                >--+-------+------->

                max=4     avg=2.33

                        (i)

in which the maximum number of exchanges drops from 5 to 4. The worst-case input
occurs when reordering (3,4,1,2) and in this case all save but the final link
are active. A breakdown of total swap counts appears below:

                            Histogram of Exchanges

     exchanges  0       1       2       3       4       5
                -       -       -       -       -       -

    (g) "rake"  1       5       8       6       3       1(*)

    (i) "comb"  1       4       8       8       3       0

                                                          * - input  4 2 3 1

Note that the reappearance of the central link (2,3) in (i) implies that a
"needless" i.e. redundant comparison might take place, as when presented the
input (1,3,2,4) requiring merely one central transposition. This cannot be
helped; the network is statistically optimal. In the extreme case, sorted data
requiring no exchanges will nonetheless compare at all link positions, including
"redundant" links -- such is the topology of CE networks. Attempts to mitigate
this, such as (software) early termination as when reaching the threshold of
maximum exchange are not cost effective: they must both count and test in order
to detect this condition at a cost often greater than a few additional link
comparisons.

The above (i) also sheds light into the subtleties of conditional probability.
While there are 4!= 24 possible inputs permutations, the first four links cannot
each successively half the size of the solution space: 24 is not 4x divisible by
2. While links 3 and 4 both execute independently with 50/50 exchange
statistics, they are nonetheless conditionally bound. If link 3 does execute
there is a 2/3 change that link 4 does not, and vise versa. In effect, when
"double winner" (max of all four) is found by exchanging element 2 with 3,
previous losers to this element may be ascribed less demotion; they are more apt
to retain a superior position and not trigger a link 4 exchange. Intuition
aside, careful analysis of exchange statistics bears this all out.


Software Implementation

Reduction in total exchanges without the use of conditional directives is the
greatest benefit of CONEX techniques. In the case of 2-ops and 3-ops a simple
software addition can reduce the explicit cost of an unnecessary comparison:

  #define sort2(a,b)   if ((a) < (b)) { (a) ^= (b); (b) ^= (a); (a) ^= (b) }
  #define op2(a,b,c)   if (sort2(a,b)) else sort2(b,c)
  #define sort3(a,b,c) sort2(a,c); op2(a,b,c)

in which the "else" elides a non-interfering CE element. The form shown above
provides exchanges without resort to additional variables / registers. If we
allow their inclusion the difference implicit in the first comparison can be
retained to speed the body of the exchange, as shown elsewhere [PaehGems]:

  #define sort2(a,b)   if ((t = (a) - (b)) < 0) { (b) += t; (a) -= t); }

With 3-ops, an exchange of the central element will elide possible swaps by
the outlying "wings"

  #define op3(a,b,c,d)   if (sort2(b,c)) else { sort2(a,b); sort2(c,d); }

These macros expand as inline code, removing subroutine overhead and form the
basic primitives used in software implementations of NICE networks.


Higher Order Networks

For N odd, we explore methods of median finding, for N even, we consider fully
sorting networks.

Sorting N=5

An efficient method for N=5 is elusive: network sizes which are a power of two
show a high degree of symmetry, sizes 2^N+1 must accommodate the new addition.
Theory prescribes 7 comparisons, as (5! = 120) < (128 = 2^7) and is realizable
in careful practice. For CE implementations, a worst case figure of 10 links is
an upper bound, this is the triangle number T(4) enumerating all pairwise
comparisons of five elements. In practice, nine links is the minimum CE
implementation. Exhaustive searching reveals that a simple elimination of one
maximal element (four steps) followed by a sort4 (five links) gives the best
sorting statistics:

    +=====================================+
    |9  6  3.133                          |
    +-------------------------------------+
    | 1 >-+-----------+-----+------->  x  |
    | 2 >-|--+--------|--+--+-----+->  x  |
    | 3 >-|--|--+-----+--|-----+--+->  x  |
    | 4 >-|--|--|--+-----+-----+---->  x  |
    | 5 >-+--+--+--+---------------->  x  |
    +=====================================+
       1   2   3   4   5   6   7   8   9
      .50 .33 .25 .20 .33 .42 .40 .40 .30

                      (j)


Note that the final sort is style (g), not (i). While not strictly a NICE
network, a compact software implementation for N=5 appears under "Further Work".

Sorting N=6

For N=6 three-sort the elements (1,3,5) and symmetrically on (2,4,6). Note that
in each case a central 2-op occurs on element 3 (4). We now center these two
elements by again using a 2-op, but this time against the outer interval formed
by the opposing 3-sort. That is, the 2-ops (1,4,5) and (2,3,6). These leads to a
highly parallel execution model in which the central two elements are located at
(3,4) albeit unordered. By the ranking nature of CE networks, elements (1,2) and
(5,6) are also properly partitioned, albeit unsorted. Three concurrent links at
(1,2) (3,4) and (5,6) thereby conclude the sort:

      13  9  4.633
  >-+-----+-----+-----+-->        >-+-----+-----+-----+-->
  >-|--+--|--+--|--+--+-->        >-|--+--|--+--|--+--+-->
  >-|--|--+--|--|--+--+-->        >-|--|--+--|--|--+--+-->
  >-|--|--|--+--+--|--+-->        >-|--|--|--+--+--|--+-->
  >-+--|--+--|--+--|--+-->        >-|--+--|--+--|--+--+-->
  >----+-----+-----+--+-->        >-+-----+-----+-----+-->
             (                                (l)k)
  first 2: .5, last 3: .2, else .33           I    II    III  IV

A rework on the right preservers overall exchange statistics while lowering the
worst-case performance to 8 swaps. Note that both networks complete in only four
stages with central stages II and III involving all elements concurrently.

If the goal is to minimize average swaps in reduced links, these suffice, at


   13  8  4.233
   >-+-----+-----------+--+---->
   >-|--+--|-----+-----|--+---->
   >-|--|--|--+--|--+--+--+--+->
   >-|--|--+--|--+--|--+--+--+->
   >-|--+-----|-----+--|--+---->
   >-+--------+--------+--+---->
      I     II   II   IV  V  VI

                 (m)

the expense of additional stages. Finally, basic methods (symmetric min/max,
applied successively) sort six elements in 12 links -- the minimum:

  12  9  4.567
  >-+--------+-----+---------------->
  >-|--+-----|-----+--+-----+---->
  >-|--|--+--+--------|--+--+--+->
  >-|--|--+--+--------+--|--+--+->
  >-|--+-----|-----+-----+--+---->
  >-+--------+-----+------------->

                 (n)


and the user may choose a smallest total number of swaps (again, 8) at the cost
of a slightly increased average number of swaps. This trade off with mild change
in topology symmetry is seen with large networks in general.

  12  8    4.7
  >-+--------+-----+---------------->
  >-|--+-----|-----+--+-----+------->
  >-|--|--+--+--------|--+--+-----+->
  >-|--+--|--+--------|--+-----+--+->
  >-|-----+--|-----+--+--------+---->
  >-+--------+-----+---------------->

                 (o)

------------
N=7 - successive 3-op median (useful in qsort partitioning)
N=8 - return to descending methods
\end{Verbatim}
\end{appendices}

\end{document}